\documentclass[prb,twocolumn,aps,showpacs,eqsecnum,amsmath]{revtex4}
\usepackage{graphicx,bm,amsmath,amssymb,times}
\begin{document}

\title{
Magnetic response and quantum critical behavior 
 in the doped two-leg extended Hubbard ladder
}

\author{M.\ Tsuchiizu and Y.\ Suzumura}

\affiliation{
Department of Physics, Nagoya University, Nagoya 464-8602, Japan}

\date{1 June, 2004}

\begin{abstract}
We have investigated quantum critical behavior 
  in the doped two-leg  extended Hubbard ladder,
     by using a weak-coupling bosonization method.
In the  ground state,  
  the dominant fluctuation changes from 
    the conventional $d$-wave-like superconducting
   (SC$d$) state   into  density-wave states,
  with increasing nearest-neighbor
 repulsions and/or decreasing doping rate. 
 The  competition  between  the SC$d$  state and 
 the charge-density-wave state coexisting with 
  the $p$-density-wave state becomes noticeable  
    on the critical  point, 
    at which  the  gap for magnetic excitations vanishes.
Based on the Majorana-fermion description of the effective theory,
  we calculate the temperature dependence of the magnetic response such as
  the spin susceptibility 
  and the NMR relaxation rate,
 which exhibit unusual properties due to two kinds 
   of spin excitation modes.
On the quantum critical point,
  the spin susceptibility
 shows paramagnetic behavior
  with logarithmic corrections and 
  the NMR relaxation rate
  also  exhibits  anomalous power-law behavior.
We discuss the commensurability effect due to the umklapp scattering
  and relevance to the two-leg ladder compounds
  Sr$_{14-x}$Ca$_x$Cu$_{24}$O$_{41}$.
\end{abstract}

 
\maketitle

\section{Introduction}

The two-leg ladder compounds, Sr$_{14-x}$Ca$_x$Cu$_{24}$O$_{41}$, 
  which exhibit the superconducting state \cite{Uehara} 
   under pressure for $x\gtrsim 12$,
 have been studied intensively
since superconductivity was confirmed in a system with ladder structure.
  \cite{Dagotto,Dagotto1999,Azuma,Ishida,Kojima}
This material, consisting of chain and ladder layers, 
  already has  holes which are doped on the ladder layer
   even for the parent material, $x=0$.
Reflecting its ladder structure, 
 there exists
   a large gap in the magnetic excitations. \cite{Imai,Takigawa}
It is known that Ca substitution yields 
an increase  of the doping rate 
  from $\delta \approx 0.07$ ($x=0$) to $\delta \approx 0.25$ ($x=12$),
\cite{Osafune} and 
  the material becomes favorable for the superconducting state.
A recent NMR measurement on superconducting 
  materials under high pressure reveals evidence for a large spin gap 
  ($\approx 200~\mathrm{K}$),
  \cite{Mayaffre,Piskunov,Fujiwara} which is much higher than the
  superconducting transition temperature $T_c$. 
From the measurement of the
 $^{63}$Cu NMR relaxation rate $T_1^{-1}$ at ladder sites,
Fujiwara \textit{et al.}
\cite{Fujiwara}
 suggested that there are two excitation modes in the normal state:
 The one at higher temperature is an activation-type mode 
 due to the spin gap 
  and the other at lower temperature is an anomalous paramagnetic mode.
In addition to the superconducting state,
  intensive studies have been devoted to
  the charge-density-wave (CDW) state,
  which is found in the  parent system ($x=0$)
  from optical measurements.
\cite{Kitano,Blumberg,Gorshunov}
Further 
the global phase diagram for overall hole doping 
\cite{Motoyama,Vuletic} shows that, with increasing the hole doping,
  the CDW state is suppressed and disappears at 
$x \simeq 9$, while the superconducting state emerges for $x\simeq 11$
 and under pressure.
From Raman scattering measurements, \cite{Gozar} 
 it has been suggested that 
 collective modes of the CDW state  exist even in the highly doped 
  superconducting material  $x=12$. 
Thus, it is expected that, by varying $x$ and temperature,
 the competition between the superconducting (SC) state and the CDW state
will become crucial in these  compounds, and nontrivial critical behavior 
 will  emerge in the competing region.

Theoretical approaches to  doped two-leg ladder systems have been performed 
  by using various kinds of methods.
It has been established that the $d$-wave-like superconducting
  (SC$d$) state becomes the most
dominant fluctuation in the ground state of the doped 
  two-leg Hubbard ladder and
$t$-$J$ ladder systems.
 \cite{Dagotto1992,Finkelstein,Gopalan,Fabrizio,Sigrist,Tsunetsugu1994,%
   Khveshchenko1994,Noack,Nagaosa,Schulz1996,Balents1996,Orignac}
When the model is extended to include parameters  of several 
 intersite Coulomb repulsions,
 other fluctuations can overcome the SC$d$ state.
A global phase diagram 
  obtained from
  a weak-coupling $g$-ology approach
  shows that
  the CDW state, $s$-wave-singlet state, and 
  $d$-density-wave states 
  can also become quasi-long-range-ordered states in  parameter space.
  \cite{Schulz1996,Orignac,Fradkin2002}
The CDW state has been confirmed  by studying numerically 
 the Hubbard model with  nearest-neighbor repulsions.
  \cite{Vojta}
A  richer phase diagram is obtained for the system at half filling
   \cite{Tsuchiizu2002b,Fradkin2002}
   and at quarter filling. \cite{Orignac2003}
Despite the huge number of theoretical works on the ordered state, 
  the critical behavior expected close to the boundary
  between different phases  
  is still unknown.
In addition, the temperature dependence of 
 magnetic quantities, such as the uniform spin susceptibility
  and the NMR relaxation rate, has been examined mainly 
 on undoped spin ladder systems and the case 
  with finite doping is not yet clarified.

In the present paper, we investigate electronic states 
  both at zero temperature and at finite temperature for the
  two-leg extended Hubbard model with finite doping,
  where special attention is focused on states 
near the phase boundary between the SC$d$ state and the 
 density-wave state.
Based on  the Majorana-fermion description,
  which is used for the low-energy effective theory
  of the spin degrees of freedom, 
we demonstrate the unconventional temperature dependence of 
both the spin susceptibility and  the NMR relaxation rate
close to the quantum critical region
 and also on the critical point,
 including logarithmic corrections.

This paper is organized as follows. 
In Sec.\ \ref{sec:model}, 
we introduce a model Hamiltonian for doped ladder systems
 and derive an effective theory describing low-energy physics, 
 using weak-coupling $g$-ology
supplemented  by bosonization and refermionization.
In Sec.\ \ref{sec:groundstate}, we investigate the ground-state phase
diagram by using the renormalization-group (RG) method, and 
clarify that the system exhibits a quantum critical behavior on the
boundary between the 
SC$d$ state and the coexisting state of CDW and $p$-density wave.
In Sec.\ \ref{sec:mag_res}, the temperature dependence of both the
 spin susceptibility and the NMR relaxation rate is examined to clarify
 their anomalous behavior in the proximity to the
 quantum critical point. 
Finally in Sec.\ \ref{sec:conclusions}, we give a summary and discuss 
  the commensurability effect due to  umklapp scattering
  and the relevance of the present calculation to the experimental results 
 for Sr$_{14-x}$Ca$_{x}$Cu$_{24}$O$_{41}$.

\section{Model Hamiltonian and Formulation}\label{sec:model}

We consider a model of a doped two-leg Hubbard ladder
   with on-site and intersite Coulomb repulsions.
The Hamiltonian is given by
\begin{equation}
H = H_0 + H_{\mathrm{int}}.
\label{eq:H}
\end{equation}
The first term $H_0$ describes electron hopping along and between the legs:
\begin{eqnarray}
H_{0} \!\! &=& \!\! - t_\parallel \sum_{j,\sigma,l}
   (c_{j,l,\sigma}^\dagger \, c_{j+1,l,\sigma}^{}+\mathrm{H.c.})
\nonumber \\
&& {} \!\! - t_\perp \sum_{j,\sigma}
   (c_{j,1,\sigma}^\dagger \, c_{j,2,\sigma}^{}+\mathrm{H.c.})
,
\end{eqnarray}
where $c_{j,l,\sigma}$ annihilates an electron of spin 
  $\sigma$ $(=\uparrow,\downarrow)$ on the $j$th rung and $l$th leg
  with $l=1,2$.
The Hamiltonian $H_{\mathrm{int}}$ represents interactions between electrons:
\begin{eqnarray}
H_{\mathrm{int}} &=&
U \sum_{j,l} n_{j,l,\uparrow} \, n_{j,l,\downarrow}
+V_\parallel \sum_{j,l} n_{j,l} \, n_{j+1,l}
\nonumber \\&& {}
+V_\perp \sum_{j} n_{j,1} \, n_{j,2},
\end{eqnarray}
where $U$ $(>0)$, $V_\parallel$ $(\ge 0)$, and $V_\perp$ $(\ge 0)$
  are coupling constants for 
  the on-site repulsion,
  the nearest-neighbor repulsion on respective chains, and
  the nearest-neighbor repulsion on a rung, respectively.
The density operators are 
   $n_{j,l,\sigma}=c_{j,l,\sigma}^\dagger \, c_{j,l,\sigma}^{}$
   and $n_{j,l}=n_{j,l,\uparrow}+n_{j,l,\downarrow}$.

Considering the two-particle interactions $H_{\mathrm{int}}$ 
  as a weak perturbation, we first 
diagonalize the single-particle hopping part $H_0$.
The diagonalization can be performed in terms of the
 the Fourier transform, 
   $c_\sigma(\bm{k})=(2N)^{-1/2} \sum_{j,l} 
   e^{-ik_\shortparallel j-ik_\perp l} c_{j,l,\sigma}$
where $\bm{k}=(k_\parallel,k_\perp)$ with $k_\perp=0,\pi$, and
the lattice spacing $a$ is set equal to 1.
Then $H_0$ is rewritten as
 $H_0 = \sum_{\bm{k},\sigma} \varepsilon(\bm{k}) \, 
  c_{\sigma}^\dagger (\bm{k}) \,   c_{\sigma} (\bm{k})$,  where
\begin{equation}
   \varepsilon(\bm{k}) = -2 t_\parallel \cos k_\parallel
       - t_\perp \cos k_\perp.
\label{dispersion}
\end{equation}
Here we consider the hole doping $\delta$ for 
 $t_\perp< 2t_\parallel \cos^2 (\pi \delta /2)$ where
    both the bonding ($k_\perp=0$) and
   antibonding ($k_\perp=\pi$) energy bands are partially filled.
In this case, 
the Fermi points are located at $k_\parallel=\pm k_{F,0}$ and 
  $\pm k_{F,\pi}$ for the bonding and antibonding bands, respectively,  
where
\begin{equation}
   k_{F,0}=\frac{\pi}{2}(1-\delta)+\lambda, \quad
   k_{F,\pi}=\frac{\pi}{2}(1-\delta)-\lambda,
\end{equation}
  and the quantity $\lambda$ is given by
\begin{equation}
\lambda \equiv 
   \sin^{-1} \left[t_\perp/\left(2t_\parallel\cos \frac{\pi}{2}\delta 
\right)\right].
\label{eq:lambda}
\end{equation}
The Fermi velocity of the bonding band and that of the antibonding band
are given by 
   $v_{F,0}=2t_\parallel \cos(\pi\delta/2 - \lambda)$
   and 
   $v_{F,\pi}=2t_\parallel \cos(\pi\delta/2 + \lambda)$,
  respectively.
For $\delta=0$, the Fermi velocity takes the common value
   $v_{F,0}=v_{F,\pi} = v_F$, where
$v_F= 2t_\parallel \left[ 1-( t_\perp/2t_\parallel )^2 \right]^{1/2}$
   for arbitrary $t_\perp$ $(<2t_\parallel)$.
In the following,
  the  difference between $v_{F,0}$ and $v_{F,\pi}$ is not taken into 
  account since we restrict ourselves to
   the small-doping case $|\delta| \ll 1$.

Let us define order parameters of possible states.
The most favorable state in doped ladders is 
the SC$d$ state  whose order parameter is 
  given by \cite{Tsuchiizu2002b}
\begin{equation}
O_{\mathrm{SC}d} = \sum_j
  \left( c_{j,1,\uparrow} c_{j,2,\downarrow}
   - c_{j,1,\downarrow} c_{j,2,\uparrow} \right).
\end{equation}
Other possible ground states are density-wave states with 
 different angular momenta.\cite{Tsuchiizu2002b,Fradkin2002}
In this paper we consider $s$-density-wave and $p$-density-wave (PDW) states.
The $s$-density-wave state is nothing but the conventional CDW state.
The order parameters of the density-wave states are given by
\begin{equation}
O_A
=
\sum_{\bm{k},\sigma, \pm}
f_A (\bm{k}) \, c_\sigma^\dagger (\bm{k}) \, c_\sigma^{} (\bm{k\pm Q})
,
\end{equation}
  where $A$ is CDW or PDW, and $\bm{Q}=(2k_F,\pi)$ 
  with $k_F=\frac{1}{2}\pi (1-\delta)$.
The form factors are given by
  $f_{\mathrm{CDW}}=1$ and $f_{\mathrm{PDW}}=i\sin k_\parallel$.
These order parameters in real space are given by
\begin{subequations}
\begin{eqnarray}
O_{\mathrm{CDW}} &=&
2\sum_{j,\sigma}
\left( c_{j,1,\sigma}^\dagger c_{j,1,\sigma}^{}
    - c_{j,2,\sigma}^\dagger c_{j,2,\sigma}^{} \right) \cos 2k_F R_j,
\nonumber \\
\\
O_{\mathrm{PDW}}&=&
\sum_{j,\sigma} 
\left( c_{j-1,1,\sigma}^\dagger c_{j,1,\sigma}^{}
    - c_{j-1,2,\sigma}^\dagger c_{j,2,\sigma}^{}
\right.\ \nonumber \\ && {} \quad \left.
- c_{j,1,\sigma}^\dagger c_{j+1,1,\sigma}^{}
    + c_{j,2,\sigma}^\dagger c_{j+1,2,\sigma}^{} \right) \cos 2k_F R_j,
\nonumber \\
\end{eqnarray}
\end{subequations}
  where $R_j=ja$.
The order parameter of the PDW state corresponds to
  that of  the spin-Peierls state in the limit of $\delta\to 0$.
One can also consider other density-wave states such as
  the  $d$-density-wave state, the $f$-density-wave state, or 
  another superconducting state with $s$-wave symmetry, 
  which are known to span the finite parameter space of 
  the ground-state phase diagram
  in the extended Hubbard ladder. \cite{Tsuchiizu2002b,Fradkin2002}
However, for the case of $U \gg V_\parallel$, $V_\perp >0$, 
  these unconventional states do not become dominant.

\subsection{$g$-ology}

Following the standard weak-coupling approach ($g$-ology),
 we linearize the energy bands around the Fermi points.
The linearized kinetic energy is given by
\begin{equation}
H_0 = \sum_{\bm{k},p,\sigma}
v_F (pk_\parallel-k_{F,k_\perp})  \,
  c_{p,\sigma}^\dagger (\bm{k}) \, c^{}_{p,\sigma} (\bm{k}),
\end{equation}
   where the index $p= +\, (-)$ denotes the right- (left-)moving electron.
We introduce field operators of the right- and left-going
electrons defined by
  $\psi_{p,\sigma,+}(x) = L^{-1/2}\sum_{k_\parallel} 
   e^{ik_\parallel x}
   c_{p,\sigma}(k_\parallel,0)$ and
  $\psi_{p,\sigma,-}(x) = L^{-1/2} \sum_{k_\parallel} 
  e^{ik_\parallel x}
   c_{p,\sigma}(k_\parallel,\pi)$ where
   $L$ is the length of the chains: $L=Na$.

The interactions near the Fermi points are written as
 $H_\mathrm{int}=\int dx \,\, \mathcal{H}_\mathrm{int}$, where
\begin{eqnarray}
\mathcal{H}_\mathrm{int}
\!\! &=& \!\!
   \frac{1}{4}
   \sum_{p,\sigma}{\sum_{\zeta_i=\pm}}'
[  g_{1\parallel}^{\epsilon\bar\epsilon} \,
      \psi_{p,\sigma,\zeta_1}^\dagger \,
      \psi_{-p,\sigma,\zeta_2}^\dagger \,
      \psi_{p,\sigma,\zeta_4}^{} \,
      \psi_{-p,\sigma,\zeta_3}^{}
\nonumber \\&& {} 
  + g_{1\perp}^{\epsilon\bar\epsilon} \,
      \psi_{p,\sigma,\zeta_1}^\dagger \,
      \psi_{-p,\bar\sigma,\zeta_2}^\dagger \,
      \psi_{p,\bar\sigma,\zeta_4}^{} \,
      \psi_{-p,\sigma,\zeta_3}^{}
\nonumber \\&& {} 
  + g_{2\parallel}^{\epsilon\bar\epsilon} \,
      \psi_{p,\sigma,\zeta_1}^\dagger \,
      \psi_{-p,\sigma,\zeta_2}^\dagger \,
      \psi_{-p,\sigma,\zeta_4}^{} \,
      \psi_{p,\sigma,\zeta_3}^{}
\nonumber \\&& {} 
  + g_{2\perp}^{\epsilon\bar\epsilon} \,
      \psi_{p,\sigma,\zeta_1}^\dagger \,
      \psi_{-p,\bar\sigma,\zeta_2}^\dagger \,
      \psi_{-p,\bar\sigma,\zeta_4}^{} \,
      \psi_{p,\sigma,\zeta_3}^{}],
\label{eq:Hint_g-ology}
\end{eqnarray} 
   where $\bar\sigma=\uparrow(\downarrow)$ for 
   $\sigma=\downarrow(\uparrow)$ and
   $\epsilon\equiv \zeta_1\zeta_3$ and $\bar\epsilon \equiv \zeta_1\zeta_2$.
The primed summation over $\zeta_i$ ($i=1,\ldots,4$) is taken
   under the condition 
   $\zeta_1 \zeta_2 \zeta_3 \zeta_4 = +1$,
   which comes from the momentum conservation 
   in the transverse direction. 
Each $g^{\epsilon\bar\epsilon}$ has two different processes
  [e.g., for $g^{++}$ 
    one has $(\zeta_1,\zeta_2,\zeta_3,\zeta_4)=(+,+,+,+)$ and $(-,-,-,-)$],
   however; these two different processes
   are given in the same form in the bosonized 
   Hamiltonian and contribute to physical quantities
   in the same manner, as will be shown later.
Then the coupling constants $g_{i\parallel}^{\epsilon\bar\epsilon}$
 and $g_{i\perp}^{\epsilon\bar\epsilon}$ 
  are written in terms of interactions of 
  the Hamiltonian (\ref{eq:H}) as
\begin{subequations}
\begin{eqnarray}
&& g_{i\parallel}^{\epsilon\bar\epsilon}
=  l_\epsilon V_\perp + m_{i,\epsilon} V_\parallel,
\label{eq:gpara} 
\\
&& g_{i\perp}^{\epsilon\bar\epsilon}
= U + l_\epsilon  V_\perp + m_{i,\epsilon} V_\parallel,
\end{eqnarray}%
\label{eq:g}%
\end{subequations} 
   where   
   $l_\pm = \pm 1$,
   $m_{1,+}=-2\cos\pi\delta \, \cos 2\lambda$,
   $m_{1,-}=-2\cos\pi\delta$,
   $m_{2,+}=+2$, and
   $m_{2,-}=+2\cos 2\lambda$.

\subsection{Bosonization}

Here we apply the Abelian bosonization method.\cite{Emery,Solyom,Gogolin_book}
The field operators of the right- and left-moving electrons 
are then written as
\begin{equation}
\psi_{p,\sigma,\zeta}(x) =
 \frac{\eta_{\sigma,\zeta}}{\sqrt{2\pi a} }
\exp \left[  ipk_{F,k_\perp} x 
 + i p\, \varphi _{p,s,\zeta}(x) \right],
\label{eq:field} 
\end{equation}
   where $s=+$ $(-)$ for $\sigma=\, \uparrow$ $(\downarrow)$.
The chiral bosons obey the commutation relations
   $[\varphi_{p,s,\zeta}(x),\varphi_{p,s',\zeta'}(x')]$
    $= ip\pi \, \mathrm{sgn}(x-x') \, 
    \delta_{s,s'}\,\delta_{\zeta,\zeta'}$
   and
   $[\varphi_{+,s,\zeta},\varphi_{-,s',\zeta'}]
    = i\pi \,\delta_{s,s'}\,\delta_{\zeta,\zeta'}$.
The Klein factors $\eta_{\sigma,\zeta}$ 
   are introduced in order to retain the correct anticommutation
   relations of the field operators between different spin and 
   band indices.
To relate the bosonic field $\varphi$ to the physical quantity, 
  we introduce  a new set of bosonic fields $\phi_{\rho\pm}$ and 
  $\phi_{\sigma\pm}$ by
\begin{subequations}
\begin{eqnarray}
\phi_{\rho +}^p(x) &=&
  \frac{1}{4} \sum_{s,\zeta=\pm}
    \varphi_{p,s,\zeta} (x)
,\\
\phi_{\rho -}^p(x) &=&
  \frac{1}{4} \sum_{s,\zeta=\pm} \zeta \,
    \varphi_{p,s,\zeta} (x)
,\\
\phi_{\sigma +}^p(x) &=&
  \frac{1}{4} \sum_{s,\zeta=\pm} s \,
    \varphi_{p,s,\zeta} (x)
,\\
\phi_{\sigma -}^p(x) &=&
  \frac{1}{4} \sum_{s,\zeta=\pm} s\zeta  \,
    \varphi_{p,s,\zeta}(x) .
\end{eqnarray}%
\end{subequations}
The phases $\phi_{\rho\pm}$ and $\phi_{\sigma\pm}$ represent charge 
   and spin fluctuations, respectively, and the suffix $+$ $(-)$ 
   refers to the even (odd) sector.

In  terms of bosonic fields, we can rewrite the kinetic energy  as
$H_0=\int dx \,\, \mathcal{H}_0$, where
\begin{equation}
{\cal H}_0
= \frac{v_F}{\pi}
\sum_{\nu=\rho,\sigma}
 \sum_{r=\pm}
\left[
  \left(  \partial_x \phi_{\nu r}^+ \right)^2
+ \left(  \partial_x \phi_{\nu r}^- \right)^2
\right]
.
\label{eq:H_harmonic}
\end{equation}
We also introduce
 the field $\phi_{\nu r}$ and its dual field
 $\theta_{\nu r}$
   defined by
  $\phi_{\nu r}=\phi_{\nu r}^+ + \phi_{\nu r}^-$ and
  $\theta_{\nu r}=\phi_{\nu r}^+ - \phi_{\nu r}^-$.
These fields satisfy the commutation relation
   $[ \phi_{\nu r}(x), \theta_{\nu' r'}(x') ] =
   -i \pi \Theta(-x+x')\delta_{r,r'}$,
   where $\Theta(x)$ is the Heaviside step function.

In terms of these bosonic fields, the order parameters are expressed as 
  $O_A = \int dx \,\, \mathcal{O}_A$, where
\begin{subequations}
\begin{eqnarray}
\mathcal{O}_{\mathrm{SC}d}(x)
&=&
\sum_{p,\zeta} \zeta \,\,
\psi_{p,\uparrow,\zeta} (x) \, 
\psi_{-p,\downarrow,\zeta} (x)
\nonumber \\
&\propto&
  e^{i \theta_{\rho+}}
  \cos \theta_{\rho-} \,
  \cos \phi_{\sigma+} \,
  \cos \phi_{\sigma-}
\nonumber \\ && 
{}  - i \,
  e^{i \theta_{\rho+}}
  \sin \theta_{\rho-} \,
  \sin \phi_{\sigma+} \,
  \sin \phi_{\sigma-} 
,
\label{eq:order_SCd}
\\ 
\mathcal{O}_{\mathrm{CDW}}(x)
&=&
\sum_{p,\sigma,\zeta}
\psi_{p,\sigma,\zeta}^\dagger (x) \, 
\psi_{-p,\sigma,-\zeta} (x) \, e^{ip2k_F x}
\nonumber \\
&\propto&
  \sin \phi_{\rho+} \,
  \cos \theta_{\rho-} \,
  \sin \phi_{\sigma+} \,
  \sin \theta_{\sigma-} 
\nonumber \\ && 
{} -
  \cos \phi_{\rho+} \,
  \sin \theta_{\rho-} \,
  \cos \phi_{\sigma+} \,
  \cos \theta_{\sigma-} ,
\label{eq:order_CDW}
\\
\mathcal{O}_{\mathrm{PDW}}(x)
&=&
\sum_{p,\sigma,\zeta} (ip) \,
\psi_{p,\sigma,\zeta}^\dagger (x) \, 
\psi_{-p,\sigma,-\zeta} (x) \, e^{ip2k_F x}
\nonumber \\
&\propto&
  \cos \phi_{\rho+} \,
  \cos \theta_{\rho-} \,
  \sin \phi_{\sigma+} \,
  \sin \theta_{\sigma-} 
\nonumber \\ && 
{} +
  \sin \phi_{\rho+} \,
  \sin \theta_{\rho-} \,
  \cos \phi_{\sigma+} \,
  \cos \theta_{\sigma-} 
.\qquad\quad
\label{eq:order_PDW}
\end{eqnarray}
\label{order-parameters}
\end{subequations}
In the present paper we do not consider the $4k_F$ CDW order parameter.
The possibility of the $4k_F$ CDW state will be discussed later in Sec.\ 
\ref{sec:groundstate}.

\begin{table*}
\caption{Position of phase locking and signs for the fixed-point 
  coupling constants, which  is essentially the same as Ref.\ 
  \onlinecite{Fradkin2002}.
The $*$ symbol indicates that a bosonic field is not locked.
   $I_i$s are integers.
}
\label{table:phase-locking}
\begin{ruledtabular}
\begin{tabular}{lccccc}
 & $\langle\theta_{\rho-}\rangle$ &
 $\langle\phi_{\sigma+}\rangle$ & $\langle\phi_{\sigma-}\rangle$ &
 $\langle\theta_{\sigma-}\rangle$ 
 & 
 $(g_{\overline{c-},s+}^*,g_{\overline{c-},s-}^*,
   g_{\overline{c-},\overline{s-}}^*,g_{s+,s-}^*,
   g_{s+,\overline{s-}}^*)$
\\
\hline
CDW+PDW & $(\pi/2)(I_0+1)+\pi I_1$ &
 $(\pi/2)I_0+\pi I_2$ & $*$ & $(\pi/2)I_0+\pi I_3$ &
 $(+,0,+,0,-)$
\\
SC$d$ & $(\pi/2)I_0+\pi I_1$ & 
 $(\pi/2)I_0+\pi I_2$ & $(\pi/2)I_0+\pi I_3$ & $*$ & 
 $(-,-,0,-,0)$
\\
\end{tabular}
\end{ruledtabular}
\end{table*}

To obtain the bosonized Hamiltonian, 
 Eq.\ (\ref{eq:field}) is substituted for 
 the interaction term Eq.\ (\ref{eq:Hint_g-ology}).
The phase field $\phi_{\rho-}$ appears
   in the form $\cos (2\phi_{\rho-}+4\lambda x)$
  where $\lambda$ is given by Eq.\ (\ref{eq:lambda}).
   We can safely assume that $t_\perp$ is relevant for $t_\perp$ being
   not very small.
In this case
  we can discard 
   the $\cos (2\phi_{\rho-}+4\lambda x)$ terms
   which become irrelevant.
We also neglect the  $\cos 2\phi_{\sigma-}  \cos 2\theta_{\sigma-}$ term, 
  because this cannot become relevant due to the scaling dimension of
   $\cos 2\phi_{\sigma-} \, \cos 2\theta_{\sigma-}$ being equal to or 
  larger than 2,
   i.e.,  this term is either marginal or irrelevant.
Then our Hamiltonian reduces to
\begin{eqnarray}
{\cal H}
\!\! &=& \!\! \frac{v_F}{\pi} 
 \sum_{r=\pm}
\left[ \sum_{p=\pm}  \left(  \partial_x \phi_{\rho r}^p \right)^2
+ \frac{g_{\rho r}}{\pi v_F} 
   (\partial_x \phi_{\rho r}^+ )
   (\partial_x \phi_{\rho r}^- ) \right]
\nonumber \\ && \!\! {}
+ \frac{v_F}{\pi}
 \sum_{r=\pm}
\left[ \sum_{p=\pm}  \left(  \partial_x \phi_{\sigma r}^p \right)^2
- \frac{g_{\sigma r}}{\pi v_F} 
   \left(\partial_x \phi_{\sigma r}^+ \right)
   \left(\partial_x \phi_{\sigma r}^- \right) \right]
\nonumber \\
&&\!\! {} 
+ \frac{1}{2\pi^2 a^2}\Bigl[ 
  g_{\overline{c-},s+} \,
    \cos 2 \theta_{\rho-} \,  \cos 2 \phi_{\sigma+}
  \nonumber \\
&& {} \qquad\qquad
+   g_{\overline{c-},s-}\,
    \cos 2 \theta_{\rho-} \, \cos 2 \phi_{\sigma-} 
  \nonumber \\
&& {} \qquad\qquad
+   g_{\overline{c-},\overline{s-}}\,
    \cos 2 \theta_{\rho-} \,\, \cos 2 \theta_{\sigma-}
\nonumber \\
&& {} \qquad\qquad
+   g_{s+, s-}\,
    \cos 2 \phi_{\sigma+} \, \cos 2 \phi_{\sigma-} \nonumber \\
&& {} \qquad\qquad
+ g_{s+,\overline{s-}}\,
  \cos 2 \phi_{\sigma+}\, \cos 2 \theta_{\sigma-}  \Bigr].
\label{eq:H_boson}
\end{eqnarray}
The coupling constants 
   for the bilinear terms of the density operators are given by
\begin{subequations}
\begin{eqnarray}
&&
g_{\rho+} =
+\frac{1}{2} \sum_{\epsilon=\pm}
   ( g_{2\parallel}^{+\epsilon}+g_{2\perp}^{+\epsilon}
    -g_{1\parallel}^{\epsilon\epsilon}), \\
&&
g_{\rho-} =
+\frac{1}{2} \sum_{\epsilon=\pm} \epsilon
   ( g_{2\parallel}^{+\epsilon}+g_{2\perp}^{+\epsilon}
    -g_{1\parallel}^{\epsilon\epsilon}), \\
&&
g_{\sigma+} =
-\frac{1}{2} \sum_{\epsilon=\pm}
   ( g_{2\parallel}^{+\epsilon}-g_{2\perp}^{+\epsilon}
    -g_{1\parallel}^{\epsilon\epsilon}), \\
&&
g_{\sigma-} =
-\frac{1}{2} \sum_{\epsilon=\pm} \epsilon
   ( g_{2\parallel}^{+\epsilon}-g_{2\perp}^{+\epsilon}
    -g_{1\parallel}^{\epsilon\epsilon}),
\end{eqnarray}%
\label{eq:grho_gsigma}%
\end{subequations}
   and the coupling constants for
   the nonlinear terms are given by
\begin{subequations}
\begin{eqnarray}
&&
g_{\overline{c-},s+}= - g_{1\perp}^{-+}, \\
&&
g_{\overline{c-},s-}= - g_{2\perp}^{-+}, \\
&&
g_{\overline{c-},\overline{s-}}=
    + g_{2\parallel}^{-+} - g_{1\parallel}^{-+}, \\
&&
g_{s+,s-} = + g_{1\perp}^{++}, \\
&&
g_{s+,\overline{s-}} = + g_{1\perp}^{--} .
\end{eqnarray}
\label{eq:g_nonlinear}%
\end{subequations}
Since there is no cosine potential for the phase
  $\phi_{\rho+}$ in the Hamiltonian (\ref{eq:H_boson}), 
  the phase $\phi_{\rho+}$ is not locked 
 even in the low-energy limit.
Then 
  the CDW state and the PDW state cannot be distinguished.
Actually, 
  by the  translation of the phase  $\phi_{\rho+} \to \phi_{\rho+}+\pi/2$,
  the order parameters 
  $\mathcal{O}_{\mathrm{CDW}}$ [Eq.\ (\ref{eq:order_CDW})] and 
  $\mathcal{O}_{\mathrm{PDW}}$ [Eq.\ (\ref{eq:order_PDW})] 
  are interchanged, while the Hamiltonian 
  (\ref{eq:H_boson}) is invariant.
Thus if the CDW state becomes (quasi-)long-range-ordered, the 
  PDW state also becomes so, and then 
  these two states coexist: We call this coexisting state 
  the CDW+PDW state.

From Eqs.\ (\ref{order-parameters}) and (\ref{eq:H_boson}), 
  the CDW+PDW state and SC$d$ state are identified by the fixed points
  of 
  $g$s as summarized in Table \ref{table:phase-locking}.
Both states have a gap in the total spin sector $\phi_{\sigma+}$.
The first reason, attributable to the difference between the CDW+PDW
  state and the SC$d$ state, is that of the locking position for the 
  $\theta_{\rho-}$ mode and the $\phi_{\sigma+}$ mode, where 
   the solution $g_{\overline{c-},s+}^*<0$ leads to 
   $\langle \theta_{\rho-}\rangle = \langle \phi_{\sigma+}\rangle = 0$ 
   or $\pi/2$ 
   for the SC$d$ state, and the solution $g_{\overline{c-},s+}^*>0$ 
  results in 
    $\langle \theta_{\rho-}\rangle \neq \langle \phi_{\sigma+}\rangle$
   for the CDW+PDW state.
The second reason is the relevance of $\phi_{\sigma-}$ or
  $\theta_{\sigma-}$:
   The CDW+PDW state is obtained for the locking of 
  $\theta_{\sigma-}$ due to the relevant 
 $g_{\overline{c-},\overline{s-}}^*$ and 
   $g_{s+,\overline{s-}}^*$ terms, while
 the SC$d$ state is obtained for that of $\phi_{\sigma-}$
  due to the relevant 
$g_{\overline{c-},s-}^*$ and $g_{s+,s-}^*$ terms.

\subsection{Refermionization and effective theory for spin modes}

The coupling constants in Eq.\ (\ref{eq:Hint_g-ology}) are not independent
   parameters due to 
   the global spin-rotation SU(2) symmetry.
In terms of the coupling constants in Eq.\ (\ref{eq:H_boson}),
  the constraint is given by
  \cite{Tsuchiizu2002b,note}
\begin{subequations}
\begin{eqnarray}
g_{\sigma+}+g_{\sigma-} - g_{s+,s-} = 0,&&
\label{eq:su2_sigma}
\\
g_{\sigma+}-g_{\sigma-} - g_{s+,\overline{s-}} = 0,&&
\\
g_{\overline{c-},s+} - g_{\overline{c-},s-}
-g_{\overline{c-},\overline{s-}}
 = 0.&&
\label{eq:su2_c-}
\end{eqnarray}
\label{eq:su2's}%
\end{subequations}
Since the SU(2) symmetry holds in the original Hubbard Hamiltonian
  (\ref{eq:H}), the coupling constants in Eq.\ (\ref{eq:H_boson}) 
  must satisfy Eq.\ (\ref{eq:su2's}) in the course of renormalization.

To appreciate the SU(2) symmetry in the effective theory
(\ref{eq:H_boson}), we fermionize it by introducing
   spinless fermion fields $\psi_{p,r}$ ($p=\pm$ and $r=\pm$):
\begin{equation}
\psi_{\pm,r}(x) = \frac{\eta_r}{\sqrt{2\pi a}}
  \, \exp\left[ \pm i \,2\phi_{\sigma r}^{\pm}(x)\right],
\label{eq:refermion}
\end{equation}
   where the index $r=+a$ $(-)$ refers to the total (relative) degrees
   of freedom of the spin mode, and $\{\eta_r,\eta_{r'}\}=2\delta_{r,r'}$.
The density operators are given by
   $ :\! \psi_{p,\pm}^\dagger \, \psi_{p,\pm}^{} \!:  \, = 
   \partial_x  \phi_{\sigma \pm}^p/\pi $.
We then introduce the Majorana fermions $\xi^n$ ($n=1-4$) by
\begin{eqnarray}
\psi_{p,+} = \frac{1}{\sqrt{2}} ( \xi_{p}^2+i\xi_{p}^1 ),
\quad
\psi_{p,-} = \frac{1}{\sqrt{2}} ( \xi_{p}^4+i\xi_{p}^3 ).
\label{eq:Majorana_fermion}
\end{eqnarray}
These fields satisfy the anticommutation relations
  $\{\xi_p^n(x),\xi_{p'}^{n'}(x')\}  =
   \delta(x-x') \, \delta_{p,p'} \, \delta_{n,n'}$.
With the help of the SU(2) constraints (\ref{eq:su2's}),
we rewrite the effective Hamiltonian (\ref{eq:H_boson}) 
 in terms of the Majorana fermions:
\begin{eqnarray}
\mathcal{H}
\!\!&=&\!\! {}
\frac{v_F}{\pi} \sum_{r}
\left[
   \sum_p \left(\partial \phi_{\rho r}^p \right)^2 
 + \frac{g_{\rho r}}{\pi v_F} 
   \left(\partial_x \phi_{\rho r}^+ \right)
   \left(\partial_x \phi_{\rho r}^- \right)
\right]
\nonumber \\ && {}
-i\frac{v_F}{2} 
\left(
  \bm{\xi}_+ \cdot \partial_x \bm{\xi}_+
- \bm{\xi}_- \cdot \partial_x \bm{\xi}_-
\right)
-\frac{g_{\sigma+}}{2} \, 
\left(
 \bm{\xi}_+ \cdot \bm{\xi}_-
\right)^2
\nonumber \\ && {}
-i\frac{v_F}{2}
\left(
 \xi_+^4 \, \partial_x \xi_+^4
- \xi_-^4 \, \partial_x \xi_-^4
\right)
\nonumber \\ && {}
- i \frac{g_{\overline{c-},st}}{2\pi a} \,
  \cos 2\theta_{\rho-} \,\,
  \bm{\xi}_+ \cdot \bm{\xi}_-
\nonumber \\ && {}
- i \frac{g_{\overline{c-},ss}}{2\pi a} \,
  \cos 2\theta_{\rho-} \,\,
  \xi_+^4 \cdot \xi_-^4
\nonumber \\ && {}
-g_{\sigma-} \,
 \left( \bm{\xi}_+ \cdot \bm{\xi}_- \right)
 \, \xi_+^4 \, \xi_-^4
, \qquad
\label{eq:Heff}
\end{eqnarray}
where $\bm{\xi}_p=(\xi_p^1,\xi_p^2,\xi_p^3)$, and the coupling constants are
\begin{equation}
g_{\overline{c-},st}\equiv -g_{\overline{c-},s+},
\quad
g_{\overline{c-},ss}\equiv
  -g_{\overline{c-},s-}+g_{\overline{c-},\overline{s-}}.
\end{equation}
These coupling constants are given
in terms of the  Hubbard interactions as 
\begin{subequations}
\begin{eqnarray}
&& \hspace*{-.5cm}
g_{\overline{c-},st}
= +U-V_\perp -2 V_\parallel \cos \pi\delta , \\
&& \hspace*{-.5cm}
g_{\overline{c-},ss} 
= +U-V_\perp + 2V_\parallel(\cos \pi \delta + 2\cos 2\lambda), \qquad
\\
&& \hspace*{-.5cm}
g_{\rho+} = +U+2V_\perp +V_\parallel [4+\cos\pi\delta(1+\cos2\lambda)],
\\
&& \hspace*{-.5cm}
g_{\rho-} = -V_\perp - V_\parallel \cos \pi\delta (1- \cos 2\lambda),
\\
&& \hspace*{-.5cm}
g_{\sigma+} = +U - V_\parallel \cos\pi\delta (1+\cos 2\lambda ),
\\
&& \hspace*{-.5cm}
g_{\sigma-} = +V_\perp + V_\parallel \cos\pi\delta (1-\cos 2\lambda ).
\end{eqnarray}%
\label{eq:init}%
\end{subequations}
Thus the effective theory for the spin sector becomes
   O(3)$\times$Z$_2$ symmetric, i.e.,
the four Majorana fermions are grouped into a singlet $\xi^4$ 
  and a triplet $\bm{\xi}$.
We note that the O(3)$\times$Z$_2$ symmetry also appears in the
low-energy effective theory of the isotropic Heisenberg
ladder.\cite{Shelton}

\section{Phase Diagram in the Ground State}\label{sec:groundstate}

We investigate the low-energy behavior  using perturbative RG analysis.
There are six independent RG equations for the scaling of the
coupling constants under the transformation of the lattice constant
$a\to a e^{dl}$.
From Eq.\ (\ref{eq:Heff}), we obtain the RG equations
\begin{subequations}
\begin{eqnarray}    
&&
\frac{d}{dl} G_{\rho-} =
 - \frac{3}{4} G_{\overline{c-},st}^2
 - \frac{1}{4} G_{\overline{c-},ss}^2, \\ 
&&
\frac{d}{dl} G_{\sigma+} =
 - G_{\sigma+}^2 - G_{\sigma-}^2
 -\frac{1}{2} G_{\overline{c-},st}^2,
\label{eq:RG_gsigma+} \\ 
&&
\frac{d}{dl} G_{\sigma-} =  - 2 G_{\sigma+} \, G_{\sigma-}
 - \frac{1}{2} G_{\overline{c-},st} \, G_{\overline{c-}ss} , \\
&&
\frac{d}{dl} G_{\overline{c-},st} = 
  -  G_{\rho-} \, G_{\overline{c-},st}
  - 2 G_{\sigma+} \, G_{\overline{c-},st}
  - G_{\sigma-} \, G_{\overline{c-},ss} ,
\nonumber \\ && {} \\
&&
\frac{d}{dl} G_{\overline{c-},ss} =
  - G_{\rho-} \, G_{\overline{c-},ss}
  -3  G_{\sigma-} \, G_{\overline{c-},st} ,
\end{eqnarray}%
\label{eq:RG}%
\end{subequations}
and $dG_{\rho+}/dl =0$ where $G(0)=g/(2\pi v_F)$.
Note that these RG equations can also be derived directly from
  Eqs.\ (\ref{eq:H_boson}) and (\ref{eq:su2's}).
Since the coupling $G_{\rho+}$ is unchanged under renormalization,
 the total charge sector is critical and has 
gapless excitations. 
The  asymptotic behavior of these coupling constants for large $l$ is
examined by integrating the RG equations (\ref{eq:RG}) numerically 
with the initial conditions (\ref{eq:init}).
It is  easily found that, in most cases, 
all the coupling constants in Eq.\ (\ref{eq:RG}) grow under
renormalization  and become relevant at large $l$.
This fact implies that 
 the all the modes except for the $\phi_{\rho+}$ mode 
  become massive in most regions of the ground-state phase diagram.
\cite{Schulz1996,Balents1996,Emery1999}
These stable fixed points are called the ``C1S0 phase,''   
where the notation C$n$S$m$ denotes
  $n$ massless boson modes in the charge sector and 
  $m$ massless boson modes in the spin sector.\cite{Balents1996}
The characteristic energy scale corresponding the mass gap can be
  roughly estimated from $|m_a|=\Lambda \, e^{-l_a}$ where $\Lambda$ is the
  high-energy cutoff of the order of the bandwidth and $l_a$ is
  determined by using the fact 
  that the corresponding coupling constant $|G_a(l)|$ 
  becomes of the order of unity at $l=l_a$.

From Eq.\ (\ref{eq:RG}), we also find that
  there are two distinct stable fixed points 
  on the plane of 
the coupling constants for the original 
  extended Hubbard ladder model with $U$ $(>0)$, 
  $V_\parallel$ $(\ge 0)$, and $V_\perp$ $(\ge 0)$.
Actually, from Table \ref{table:phase-locking}, one obtains that
 the coupling constants $(G_{\rho-},G_{\sigma+},G_{\sigma-},
   G_{\overline{c-},st},G_{\overline{c-},ss})$ flow to
 $(-,-,+,-,+)$ for the CDW+PDW state and 
 $(-,-,-,+,+)$ for the SC$d$ state.
This means that there are two distinct phases in the ground states of
  the extended Hubbard ladder model, and that the system exhibits a quantum
  phase transition on a critical point between two phases.
In order to examine 
  critical properties in the ground state, 
  we analyze Eq.\ (\ref{eq:Heff}) in more detail 
  by deriving the effective theory for low-energy properties.

Here we assume that the mass of the charge mode of odd sector
   ($\rho -$) is larger than those of the spin modes ($\sigma\pm$),
  so that the  $\theta_{\rho-}$ fields are locked by cosine potential
  below the scale of the mass $m_{\rho-}$.
This assumption will be examined later.
The effective low-energy theory is obtained from Eq.\ (\ref{eq:H_boson})
  by taking an average:
\begin{equation}
c_{\overline{\rho-}}\equiv\langle\cos2\theta_{\rho-}\rangle.
\end{equation}
Then we have
\begin{eqnarray}
\mathcal{H}_\sigma
\!\!&=&\!\! {}
-i\frac{v_F}{2}  
\left( \bm{\xi}_+ \cdot \partial_x \bm{\xi}_+
     - \bm{\xi}_- \cdot \partial_x \bm{\xi}_-  \right)
 - i m_t^0 \, \bm{\xi}_+ \cdot \bm{\xi}_-
\nonumber \\ && {}
-i\frac{v_F}{2}
\left( \xi_+^4 \, \partial_x \xi_+^4
     - \xi_-^4 \, \partial_x \xi_-^4 \right)
- i m_s^0 \,  \xi_+^4 \, \xi_-^4
\nonumber \\ && {}
-\frac{g_{\sigma+}}{2} \, 
\left( \bm{\xi}_+ \cdot \bm{\xi}_- \right)^2
- g_{\sigma-} \left( \bm{\xi}_+ \cdot \bm{\xi}_- \right)
 \, \xi_+^4 \, \xi_-^4 , \qquad
\label{eq:Heff_spin}
\end{eqnarray}
where we have introduced
\begin{equation}
m_t^0 \equiv \frac{c_{\overline{\rho -}}}{2\pi a}  g_{\overline{c-},st},
\quad
m_s^0 \equiv \frac{c_{\overline{\rho -}}}{2\pi a} g_{\overline{c-},ss}.
\label{eq:mtms0}
\end{equation}
Such a mean-field treatment of the charge sector has also been utilized
in the context of carbon nanotubes.\cite{Tsvelik_nanotube}
We note that this low-energy effective theory  takes the same form as
that of  the isotropic Heisenberg
ladder.\cite{Shelton}
In terms of the original Hubbard interactions these masses of 
  Eq.\ (\ref{eq:mtms0}) are given by
\begin{subequations}
\begin{eqnarray}
&&
m_t^0 = \frac{c_{\overline{\rho -}}}{2\pi a}  
\left[
 U-V_\perp -2 V_\parallel \cos \pi\delta 
\right]
,
\\
&&
m_s^0 = \frac{c_{\overline{\rho -}}}{2\pi a}
\left[
U-V_\perp + 2V_\parallel (\cos \pi \delta + 2\cos 2\lambda)
\right]. \qquad
\end{eqnarray}%
\end{subequations}
Thus the magnitude of the masses can be tuned by the interactions and
doping. 
It is known that, when $m_s^0,m_t^0\ne0$, the quartic marginal terms lead
   to mass renormalization, $m_s^0 \to m_s$ and
   $m_t^0 \to m_t$, where\cite{Shelton,Gogolin_book}
\begin{subequations}
\begin{eqnarray}
&&
m_t =  m_t^0 
  - \frac{g_{\sigma+}}{\pi v_F} m_t^0 \ln \frac{\Lambda}{|m_t^0|}
  - \frac{g_{\sigma-}}{2\pi v_F} m_s^0 \ln \frac{\Lambda}{|m_s^0|},
\qquad
\label{eq:mt_tilde} \\ &&
m_s =   m_s^0
  - \frac{3g_{\sigma-}}{2\pi v_F} m_t^0 \ln \frac{\Lambda}{|m_t^0|},
\label{eq:ms_tilde}%
\end{eqnarray}%
\label{eq:mtms}%
\end{subequations}
and $\Lambda$ is a high-energy cutoff.
Then Eq.\ (\ref{eq:Heff_spin})  reduces to
\begin{eqnarray}
\mathcal{H}_\sigma
&\!\!=\!\!& {}
-i\frac{v_F}{2} 
\left( \bm{\xi}_+ \cdot \partial_x \bm{\xi}_+
     - \bm{\xi}_- \cdot \partial_x \bm{\xi}_- \right)
 - i m_t \, \bm{\xi}_+ \cdot \bm{\xi}_-
\nonumber \\ && {}
-i\frac{v_F}{2}
\left( \xi_+^4 \, \partial_x \xi_+^4
     - \xi_-^4 \, \partial_x \xi_-^4 \right)
- i m_s \,  \xi_+^4 \, \xi_-^4 .
\label{eq:Majorana_Hamiltonian}
\end{eqnarray}

Low-energy properties become more transparent by introducing
 four copies of the one-dimensional quantum Ising model:
\begin{eqnarray}
H_{\mathrm{QI}} = - \sum_j \sum_{l}
\left( J \sigma_{j,l}^z \, \sigma_{j+1,l}^z + h_l \, \sigma_{j,l}^x \right),
\label{eq:QI}
\end{eqnarray}
where  $\sigma_j^z$ and $\sigma_j^x$ are the Pauli matrices and $l=1,2,3,4$.
This model is equivalent to the Majorana-fermion
theory with the central charge $c=1/2$.
The  operator $\mu$, 
  being dual to the spin operator  $\sigma$,
 is defined as\cite{Gogolin_book}
\begin{eqnarray}
\mu^z_{j+1/2,l} = \prod_{i=1}^j \sigma_{i,l}^x, \quad
\mu^x_{j+1/2,l}= \sigma^z_{j,l} \, \sigma^z_{j+1,l},
\end{eqnarray}
which is known as the Kramers-Wannier transformation in the
  one-dimensional quantum Ising model.
These variables satisfy
  $[\sigma_{i,l}^z , \mu_{j+1/2,l}^z]=0$ for $i>j$ and
  $\{\sigma_{i,l}^z , \mu_{j+1/2,l}^z \} = 0$ for $i \le j$.
In terms of $\sigma^z$ and $\mu^z$, i.e., the Ising order and disorder 
   parameters, the Majorana fermions can be constructed as 
\begin{eqnarray}
\eta_{j,l} = \kappa_l \, \sigma_{j,l}^z \, \mu_{j-1/2,l}^z, \quad
\zeta_{j,l} = i\kappa_l \, \sigma_{j,l}^z \, \mu_{j+1/2,l}^z,
\end{eqnarray}
  where $\kappa_l$ is the Klein factor.
One can easily check the anticommutation relation of the Majorana
fermions, $\{\eta_{i,l},\eta_{j,m}\}=\{\zeta_{i,l},\zeta_{j,m}\}
=2\delta_{i,j} \delta_{l,m}$ and $\{\eta_{i,l},\zeta_{j,m}\}=0$.
By using
$\xi_+^l=(-\eta_l + \zeta_l)/\sqrt{2}$ and
$\xi_-^l=(\eta_l + \zeta_l)/\sqrt{2}$,
 and by taking the continuum limit,
 the quantum Ising Hamiltonian Eq.\ (\ref{eq:QI}) reproduces
   Eq.\ (\ref{eq:Majorana_Hamiltonian}) where
   $v_F =2J$, $m_t=2(h_1-J)$, and $m_s=2(h_4-J)$ with $h_1=h_2=h_3$.
It is well known that the  Ising model (\ref{eq:QI}) 
  exhibits a quantum critical point at $h_l=J$. \cite{Sachdev_book}
For $h_l<J$, the ordered state is obtained, i.e., the order
  parameter $\sigma_l$ has a finite expectation value.
For $h_l>J$, on the other hand,
  we have the disordered state where 
  the expectation value of $\sigma_l$ becomes zero, while
  the disorder parameter $\mu_l$ 
  has a finite expectation value.
On the critical point $h_l=J$, the corresponding mass in Eq.\ 
  (\ref{eq:Majorana_Hamiltonian}) vanishes
  with its central charge 
  $c=\frac{1}{2}$
  for each Ising chain. \cite{CFT}
Thus the ground-state properties are
  determined from the sign of masses $m_t$ and  $m_s$.
When $m_t<0$, i.e., the Ising model with $l=1,2,3$ is in the 
  ordered phase, 
  we have $\langle\sigma_1\rangle=\langle\sigma_2\rangle
   =\langle\sigma_3\rangle\neq 0$ 
  and $\langle\mu_1\rangle=\langle\mu_2\rangle=\langle\mu_3\rangle= 0$,
  and vice versa.
In the same manner, we have $\langle\sigma_4\rangle\neq 0$ and 
  $\langle\mu_4\rangle= 0$ for $m_s<0$, 
  while $\langle\sigma_4\rangle=0$ and $\langle\mu_4\rangle\neq 0$ for $m_s>0$.
In terms of the Ising variables, the order parameters
  Eq.\ (\ref{order-parameters}) are rewritten as
\begin{subequations}
\begin{eqnarray}
\mathcal{O}_{\mathrm{SC}d} \propto 
e^{i \theta_{\rho+}} \, (\mu_1 \, \mu_2 \, \mu_3) \, \mu_4 ,&& \\ 
\mathcal{O}_{\mathrm{CDW}} \propto \sin \phi_{\rho+} \, 
(\sigma_1 \, \sigma_2 \, \sigma_3) \, \mu_4, && \\
\mathcal{O}_{\mathrm{PDW}} \propto
  \cos \phi_{\rho+} \, (\sigma_1 \, \sigma_2 \, \sigma_3) \, \mu_4 .&&
\end{eqnarray}
\end{subequations}
As noted in the preceding section,
we find that both the CDW and PDW states have the same structure
  for the spin degrees of freedom, 
  since the field $\phi_{\rho+}$ is unlocked 
  due to the doping effect. 
When  $m_t<0$ and $m_s>0$, 
  the CDW+PDW state becomes quasi-long-range ordered.
In the case $m_t>0$ and $m_s>0$, the dominant fluctuation is 
  the SC$d$ state, which is  called the ``Luther-Emery liquid.''
  \cite{Ledermann,Tsuchiizu2001}
The possible ground states and those order parameters 
 are summarized in Table \ref{table:phase}.

\begin{table}[t]
\caption{
Possible phases and related quantities:
 the signs of masses ($m_t$ and $m_s$) and order parameters.
We have assumed $c_{\overline{\rho -}}
  \equiv \langle \theta_{\rho-} \rangle=0$ mod $\pi$ in Eq.\ (\ref{eq:mtms0}).
}
\label{table:phase}
\begin{ruledtabular}
\begin{tabular}{lcccc}
               & $m_t$ & $m_s$ & Order parameters\\ \hline
 CDW + PDW     &  $-$  &     $+$    &  
     $\langle\sigma_{1,2,3}\rangle\neq0,  \,\, 
      \langle\mu_4\rangle\neq0 $  \\
 SC$d$         &  $+$  &     $+$    &   
     $\langle\mu_{1,2,3}\rangle\neq0, \,\, 
      \langle\mu_4\rangle\neq0 $ 
\end{tabular}
\end{ruledtabular}
\end{table}

\begin{figure}[t]
\includegraphics[width=6.cm]{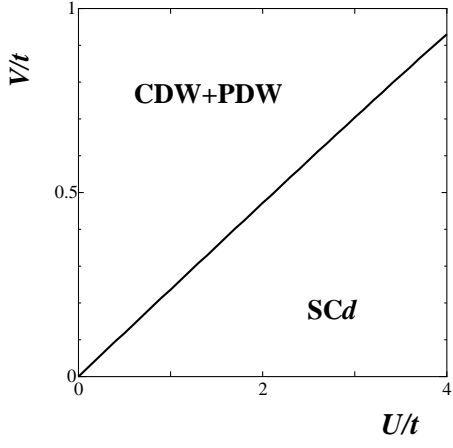}
\caption{
The ground-state phase diagram on the plane of $U/t$ and
  $V/t$ where $t \equiv t_\perp=t_\parallel$ and $\delta=0.2$.
}
\label{fig:phase_a}
\end{figure}

\begin{figure}[b]
\includegraphics[width=6.cm]{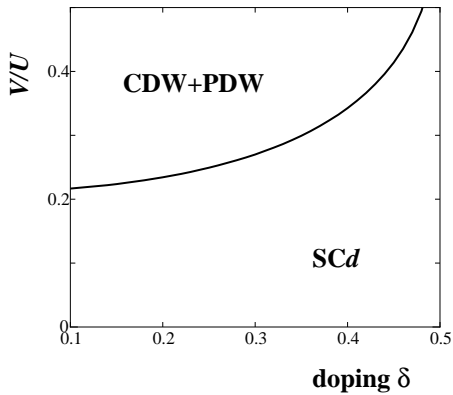}
\caption{
The ground-state phase diagram on the plane of doping rate $\delta$ and
  ratio $V/U$ where $V \equiv V_\parallel=V_\perp$ and $U/t=3$ with
   $t \equiv t_\perp=t_\parallel$.
}
\label{fig:phase_b}
\end{figure}

Let us examine the behavior in more detail using
the scaling equations for the coupling constants  in
the effective Hamiltonian (\ref{eq:Heff_spin}).
The scaling equations for the coupling constants are given by
\cite{Tsuchiizu2002b}
\begin{subequations}
\begin{eqnarray}
&&
\frac{dM_{t}}{dl} =
M_{t}-2 M_{t}G_{\sigma+}-M_{s}G_{\sigma-},
\label{eq:dG_t/dl}\\ 
&&
\frac{dM_{s}}{dl} =
M_{s} -3  M_{t}G_{\sigma-},
\label{eq:dG_s/dl}\\ 
&&
\frac{dG_{\sigma+}}{dl} =
-G_{\sigma+}^2 - G_{\sigma-}^2 - M_{t}^2,
\label{eq:dG_sigma+/dl}\\
&&
\frac{dG_{\sigma-}}{dl} =
-2G_{\sigma+}G_{\sigma-}-M_{t}M_{s},
\label{dG_sigma-/dl}%
\end{eqnarray}%
\label{eq:dG_spin}%
\end{subequations}
where $dl=da/a$, $M_{t}=m_t^0 a/v_F$,
$M_{s}=m_s^0 a/ v_F$, and
$G_{\sigma\pm}=g_{\sigma\pm}/2\pi v_F$.
The couplings $M_t$ and $M_s$ are relevant,
 while $G_{\sigma\pm}$ are marginal.
These RG equations, which are analyzed
  in a  way similar to Ref.\ \onlinecite{Tsuchiizu2002b}, have
  two kinds of stable fixed point $M_t^*=\pm\infty$. 
The ground-state phase diagram, which is summarized in Table
\ref{table:phase}, is shown in Fig.\ \ref{fig:phase_a} 
 on the plane of $U/t$ and 
  $V/t$ ($t_\parallel=t_\perp \equiv t$, 
   $V_\parallel= V_\perp\equiv V$).
The SC$d$ quasi-long-range-ordered state, which is obtained 
  for $V=0$, is destabilized by 
the intersite Coulomb repulsion and 
  changes  into the CDW+PDW state.
The ground-state phase diagram 
  on the plane of the doping $\delta$ and the ratio $V/U$
  is shown in Fig.\ \ref{fig:phase_b},
   where the SC$d$ state is stabilized due to the suppression of 
  the intersite repulsion by the doping.

Here we discuss the possibility of the $4k_F$ CDW state.
It is known that the $4k_F$ CDW state also becomes quasi-long-range ordered 
  in the whole region of the phase diagrams, Figures
  \ref{fig:phase_a} and \ref{fig:phase_b}, since the correlation
  function $4k_F$ CDW state decays \cite{Nagaosa,Schulz1996}
  as $1/r^{2K_{\rho+}}$ where $K_{\rho+}$ is the Tomonaga-Luttinger 
  parameter for the total charge sector,
  $K_{\rho+}\equiv [(2\pi v_F -g_{\rho+})/(2\pi v_F +g_{\rho+})]^{1/2}$.
Since the correlation function of the CDW and PDW states decays as
   $1/r^{K_{\rho+}/2}$, 
  we find that the $4k_F$ CDW state is still a subdominant fluctuation 
   in  the CDW+PDW state.
On the other hand, in the SC$d$ state, 
  the $4k_F$ CDW can become the dominant fluctuation
   for $K_{\rho+}<1/2$, since the exponent of the SC$d$ correlation 
  function is given by $1/2K_{\rho+}$.
However, at close to half filling, it has been confirmed that 
  the exponent $K_{\rho+}$ in ladder systems 
  reaches universal value $K_{\rho+}^* \to 1$ as $\delta \to 0$.
  \cite{Schulz1999,Konik,Siller}
Thus in the SC$d$ state of the phase diagram,
  we find that the correlation function of the SC$d$ state becomes 
  dominant  and that of the
  $4k_F$ CDW state is subdominant  close to half filling.

\begin{figure}[t]
\includegraphics[width=6.cm]{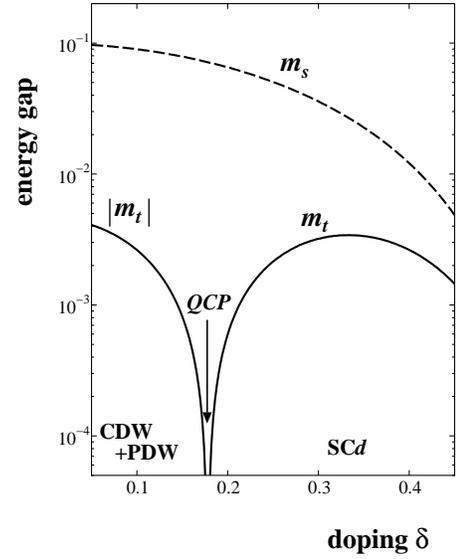}
\caption{
The doping dependence of the energy gaps 
$m_t$ and  $m_s$ with $U/t=3$, 
 $V_\parallel/t=V_\perp/t=0.7$, and $t=t_\parallel=t_\perp=1$.
The quantum critical point (QCP) 
  is at $\delta=\delta_c\approx 0.18$.
The CDW+PDW quasi-long-range-ordered state is obtained for 
  $\delta < \delta_c$, while the
the $d$-wave SC quasi-long-range-ordered state is obtained for
 $\delta>\delta_c$.
}
\label{gap}
\end{figure}

The doping dependences of the gaps,\cite{Tsuchiizu2003}
  which are roughly estimated from 
   $|m_a|=\Lambda \, e^{-l_a}$,   are shown in Fig.\ \ref{gap}.
The Majorana triplet gap $m_t$ collapses
on the boundary between the CDW+PDW state and the SC$d$ state
  where the system is strongly fluctuating
  due to  the competition between these two states.
From the perturbative RG method,
  we cannot determine the precise magnitude of the gaps; however,
  the qualitative features of the gap associated with 
  the phase transition (i.e., $m_t$ in the present case) 
  can be captured close to the quantum critical point (QCP).
The critical properties are
  described by the SU(2)$_2$ Wess-Zumino-Novikov-Witten (WZNW) model 
\cite{Shelton,Gogolin_book,Tsvelik1990,Starykh}
  and is also characterized by the C1S$\frac{3}{2}$ state.\cite{Schulz1996}
In the present analysis, we have assumed $m_{\rho-}>(m_s,|m_t|)$ 
  to derive the effective low-energy theory Eq.\ (\ref{eq:Heff_spin}).
Actually, by using  Eq.\ (\ref{eq:RG}),  we find that 
  $m_{\rho-}$ becomes the largest among the three
  $m_{\rho-}$, $m_t$,  and $m_s$.
However, $m_{\rho-}$ is not much larger than the Majorana singlet gap
  $m_s$, but is of the same order as $m_s$; 
  we find from the above rough estimation that
  $m_{\rho-}/m_s \approx 1.4 - 1.6$ for $\delta = 0.1 - 0.4$.
From the recent theoretical studies on the 
  multi component one-dimensional systems,
  it has been proposed that \cite{Azaria1998,Azaria1999,Assaraf}
   a symmetry, which is broken even in the microscopic Hamiltonian,
   can be restored nontrivially at low energies, due to coupling 
   terms between different modes.
This mechanism is called a dynamical symmetry enlargement (DSE) 
 whose possibility has also been examined
  by the nonperturbative approach. \cite{Assaraf}
Once the DSE is realized, the resultant gaps for different modes
  become identical.
Based on these theories, the present results with $m_{\rho-}\approx m_s$ 
  may suggest the occurrence of the DSE between the charge sector ($\rho-$)
  and the Majorana singlet sector where excitations form an O(3) multiplet.
However, the study of the DSE is beyond 
  the naive perturbative RG approach \cite{Azaria1998}
  and thus the DSE in the present case remains unclear.

\section{Spin Susceptibility and NMR Relaxation Rate}\label{sec:mag_res}

In this section we study the uniform spin
susceptibility and the NMR relaxation rate at finite
temperature from  two different approaches
  by extending previous calculations on the single chain
  \cite{Bourbonnais,Nelisse}
   or on the undoped Heisenberg ladder.
  \cite{Kishine,Ivanov1999,Damle,Citro}
One approach is the random-phase approximation (RPA)
 combined with the renormalization
group method and 
another is direct calculation in terms of
 the low-energy effective Hamiltonian
 (\ref{eq:Majorana_Hamiltonian}).
The former has the advantage of reproducing high-temperature behavior,
while the latter is appropriate for 
   describing  the low-temperature asymptotics.

First we introduce the spin-$\frac{1}{2}$ operator:
\begin{equation}
\bm{S} (\bm{q}) \equiv \frac{1}{2} 
\sum_{\bm{k},\sigma_1,\sigma_2}
   c_{\sigma_1}^\dagger(\bm{k}+\bm{q}) \,
   \bm{\sigma}_{\sigma_1,\sigma_2}^{} \,
   c_{\sigma_2}^{} (\bm{k}),
\end{equation}
   with $\bm{q}=(q_\parallel,q_\perp)$ and 
   the Pauli matrices $\bm{\sigma}_{\sigma_1,\sigma_2}$.
The generalized spin susceptibility is given by
\begin{equation}
\chi(\bm{q},i\omega_n)
\equiv
\frac{1}{2L} \int_0^\beta d\tau 
\left\langle  T_\tau
S^{\alpha}(\bm{q},\tau) \,S^{\alpha}(-\bm{q},0)
\right\rangle e^{i\omega_n \tau}.
\label{eq:chi}
\end{equation}
Equation (\ref{eq:chi}) is independent of $\alpha$ due to the 
   spin-rotational SU(2) symmetry,   
 where $\alpha$ stands for the orientation of the magnetic field.
The noninteracting susceptibility
  [i.e., Eq.\ (\ref{eq:chi}) without interactions] is given by
\begin{equation}
\chi_0(\bm{q},i\omega_n)
=
\frac{1}{4L}\sum_{k_\parallel,k_\perp}
 \frac{  f(\varepsilon(\bm{k}+\bm{q})) -  f(\varepsilon(\bm{k}))  }
      {i\omega_n + \varepsilon(\bm{k})-\varepsilon(\bm{k}+\bm{q})},
\label{eq:chi0general}
\end{equation}
  where $f(\varepsilon)$ is the Fermi distribution function 
  $f(\varepsilon)=1/[e^{\beta(\varepsilon-\mu)}+1]$, and
  $\omega_n$ and $\mu$ are the Matsubara frequency and  the chemical
  potential, respectively.
In the continuum limit, we can split the spin operator into a
 uniform part varying slowly in space and a staggered oscillation 
   part, as
\begin{equation}
\bm{S} (x,q_\perp) = \bm{J}_r(x) + (-1)^{x/a} \bm{n}_r(x),
\label{eq:spin_op}
\end{equation}
   where $r=+$ $(-)$ for $q_\perp=0$ $(\pi)$.
The uniform part ($q_\parallel \approx 0$) of the spin operator is given by
\begin{subequations}
\begin{eqnarray}
&&\bm{J}_+(x)= 
\frac{1}{2}\sum_{p,\zeta} \sum_{\sigma_1,\sigma_2}
\psi_{p,\sigma_1,\zeta}^\dagger(x) \,
  \bm{\sigma}_{\sigma_1,\sigma_2}^{} \,
\psi_{p,\sigma_2,\zeta}^{}(x),
\\
&&\bm{J}_-(x)= 
\frac{1}{2}\sum_{p,\zeta} \sum_{\sigma_1,\sigma_2}
\psi_{p,\sigma_1,\zeta}^\dagger(x) \,
  \bm{\sigma}_{\sigma_1,\sigma_2}^{} \,
\psi_{p,\sigma_2,-\zeta}^{}(x), \qquad\quad
\end{eqnarray}
\end{subequations}
  where $\psi_{p,\sigma,\zeta}(x)$ is given by Eq.\ (\ref{eq:field}).
By using Eqs.\ (\ref{eq:field}), (\ref{eq:refermion}), and
  (\ref{eq:Majorana_fermion}),
  the spin operator $\bm{J}_\pm(x)$ is 
  expressed in terms of the Majorana
  fermions as
\begin{subequations}
\begin{eqnarray}
&&\bm{J}_+(x) =
+\frac{i}{2} \sum_p \bm{\xi}_p(x) \times \bm{\xi}_p(x) ,
\\
&&\bm{J}_-(x) =
-i \sum_p \bm{\xi}_p(x) \, \xi_p^4(x) ,
\end{eqnarray}%
\label{eq:spin_current}%
\end{subequations}
 where $\bm{J}_\pm=(J_\pm^x,J_\pm^y,J_\pm^z)=(J_\pm^1,J_\pm^2,J_\pm^3)$.
The staggered part ($q_\parallel \approx 2k_F$ and $q_\perp=\pi$)
 of the spin operator is given by
\begin{equation}
\bm{n}_-(x)= 
\frac{(-1)^{x/a}}{2}\sum_{p,\zeta} \sum_{\sigma_1,\sigma_2}
\psi_{p,\sigma_1,\zeta}^\dagger(x) \,
  \bm{\sigma}_{\sigma_1,\sigma_2}^{} \,
\psi_{-p,\sigma_2,-\zeta}^{}(x).
\end{equation}
Here we do not consider the component with 
  $(q_\parallel,q_\perp) \approx (2 k_F,0)$,
  which would become irrelevant in the low-energy limit
  due to the relevant $t_\perp$. \cite{Tsuchiizu1997}
By using  Eq.\ (\ref{eq:field}), the operator $\bm{n}_-$ is 
  rewritten as
\begin{subequations}
\begin{eqnarray}
n^x_-(x)
\!\! &=& \!\!
\frac{-2i}{\pi a}
( \cos \tilde\phi_{\rho+}
  \cos \theta_{\rho-} \sin \theta_{\sigma+} \cos \phi_{\sigma-}
\nonumber \\ && {} \quad
- \sin \tilde\phi_{\rho+}
  \sin \theta_{\rho-} \cos \theta_{\sigma+} \sin \phi_{\sigma-}) , \qquad
\\ \nonumber \\
n^y_-(x)
\!\! &=& \!\!
\frac{-2i}{\pi a}
( \cos \tilde\phi_{\rho+}
  \cos \theta_{\rho-} \cos \theta_{\sigma+} \cos \phi_{\sigma-}
\nonumber \\ && {} \quad
+ \sin \tilde\phi_{\rho+}
  \sin \theta_{\rho-} \sin \theta_{\sigma+} \sin \phi_{\sigma-}), \qquad
\\ \nonumber \\
n^z_- (x)
\!\! &=& \!\!
\frac{2}{\pi a}
( \cos \tilde\phi_{\rho+}
  \cos \theta_{\rho-} \cos \phi_{\sigma+} \sin \theta_{\sigma-}
\nonumber \\ && {} \quad
- \sin \tilde\phi_{\rho+}
  \sin \theta_{\rho-} \sin \phi_{\sigma+} \cos \theta_{\sigma-}). \qquad
\end{eqnarray}%
\label{eq:n-_boson}%
\end{subequations}
  where $\tilde\phi_{\rho+}=\phi_{\rho+}-\pi\delta x$.
Here we note that $\bm{n}_-(x)$ expresses the incommensurate spin-density wave.

\begin{figure}[t]
\includegraphics[width=8.5cm]{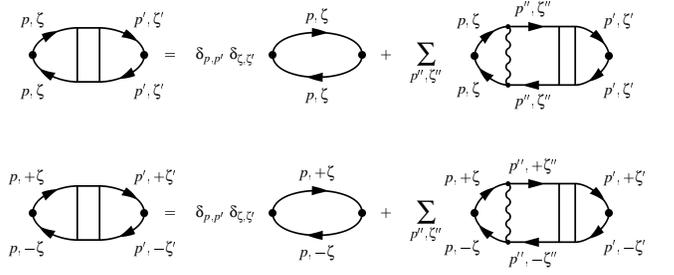}
\caption{
The spin susceptibility $\chi(q_\parallel,q_\perp,\omega)$ 
  with $q_\perp =0$ (upper diagram) and $\pi$ (lower diagram), 
  and small $q_\parallel$.
The subscript $p=+$ $(-)$ refers to right- (left-)moving electrons, while 
  $\zeta=+$ $(-)$ refers to electrons on the bonding (antibonding) band.
}
\label{fig:chi_rpa}
\end{figure}

\subsection{Uniform spin susceptibility}

Here we calculate the uniform spin susceptibility.
In terms of the spin operators $\bm{J}_\pm$,
   the spin susceptibility $\chi(\bm{q},i\omega_n)$ for small
  $q_\parallel$ is given by
\begin{equation}
\chi_{\mathrm{uni}}(\bm{q},i\omega_n)
\equiv
\frac{1}{2L}
\int_0^\beta d\tau 
\left\langle  T_\tau
J^{\alpha}_{r}(q_\parallel,\tau) \,
J^{\alpha}_{r}(-q_\parallel,0)
\right\rangle 
e^{i\omega_n \tau},
\end{equation}
  where $r=+$ $(-)$ in the right-hand side (RHS) 
  corresponds  to $q_\perp=0$ $(\pi)$ in the LHS and 
  $\alpha$ denotes the orientation of the magnetic field.
First we  calculate $\chi_{\mathrm{uni}}(\bm{q},i\omega_n)$ within the RPA 
 by using the formulation of $g$-ology (\ref{eq:Hint_g-ology}).
The magnetic susceptibility 
  $\chi(q_\parallel,q_\perp,i\omega_n)$ for small $q_\parallel$ 
  is calculated from the diagrams of the RPA given in Fig.\ \ref{fig:chi_rpa}.
We note that the same result would be derived from the 
  path integral formalism. \cite{Nelisse}
The explicit forms for the susceptibility are given  by
\begin{widetext}
\begin{subequations}
\begin{eqnarray}
&&
\chi_{\mathrm{uni}}(q_\parallel,0,i\omega_n) =
\frac{ \left[
   \chi_0^+(q_\parallel,0,i\omega_n)+\chi_0^-(q_\parallel,0,i\omega_n) 
  \right]
 + 2 (g_{1\perp}^{++} + g_{1\perp}^{--}) \,
   \chi_0^+(q_\parallel,0,i\omega_n) \, \chi_0^-(q_\parallel,0,i\omega_n)}
{ 1-(g_{1\perp}^{++} + g_{1\perp}^{--})^2 \, 
  \chi_0^+(q_\parallel,0,i\omega_n) \, \chi_0^-(q_\parallel,0,i\omega_n) },
\label{eq:chi_uni_0} \\
&&
\chi_{\mathrm{uni}}(q_\parallel,\pi,i\omega_n) =
\frac{\left[
 \chi_0^+(q_\parallel,\pi,i\omega_n)
  +\chi_0^-(q_\parallel,\pi,i\omega_n) \right]
 + 2 (g_{1\perp}^{+-} + g_{1\perp}^{-+}) \,
   \chi_0^+(q_\parallel,\pi,i\omega_n) \,
   \chi_0^-(q_\parallel,\pi,i\omega_n)}
{1-(g_{1\perp}^{+-} + g_{1\perp}^{-+})^2 \, 
  \chi_0^+(q_\parallel,\pi,i\omega_n) \,
  \chi_0^-(q_\parallel,\pi,i\omega_n)} ,
\end{eqnarray}
\end{subequations}
\end{widetext}
where $\chi_0^{p}(q_\parallel,q_\perp,i\omega_n)$ is given by 
  the noninteracting spin susceptibility per branch $p=+$ ($-$): 
\begin{equation}
\chi_0^{p}(\bm{q},i\omega_n)
\equiv
\frac{1}{4L}
\sum_{k_\parallel \approx pk_F}
\,\,
\sum_{k_\perp}
 \frac{  f(\varepsilon(\bm{k}+\bm{q})) -  f(\varepsilon(\bm{k}))  }
      {i\omega_n + \varepsilon(\bm{k})-\varepsilon(\bm{k}+\bm{q})}
.
\label{eq:chi0}
\end{equation}
It is crucial to take into account the effect of the curvature of the
  dispersion in the noninteracting susceptibility $\chi_0^p$, i.e.,
  the explicit form of $\varepsilon(\bm{k})$  
   given in Eq.\ (\ref{dispersion}) is retained in the calculation,
   to obtain a reasonable temperature dependence.

By using Eq.\ (\ref{eq:chi_uni_0}) with 
  the relations Eqs.\  (\ref{eq:g_nonlinear}) and (\ref{eq:su2's}),
  the uniform spin susceptibility 
  $\chi_s(T)=\chi_{\mathrm{uni}}(q_\parallel,0,0)|_{q_\parallel \to 0}$
  is calculated as
\begin{equation}
\chi_s(T) = \frac{\chi_0(T)}{1 -  g_{\sigma+} \chi_0(T) } ,
\end{equation}
   where 
   $\chi_0^+(0,0  )=\chi_0^-(0,0) \,\, [ \, \equiv  \! \chi_0(T)/2]$.
We note that the susceptibility $\chi_0(T)$ for $T=0$ is given
by $\chi_0(0)=(v_{F,0}+v_{F,\pi})/(4\pi v_{F,0}v_{F,\pi})$, which 
   reduces to $\chi_0(0) \to 1/(2\pi v_F)$
  in the limit of a single chain because $v_{F,0},v_{F,\pi}\to v_F$.

All the diagrams given in Fig.\ \ref{fig:chi_rpa} are non singular because
  only the small-$q_\parallel$ components are taken into account.
The effect of the one-dimensional fluctuations appears through
   logarithmic corrections to $g_{\sigma+}$.
This process can be accomplished by replacing the coupling constant 
  $g_{\sigma+}$
  with the renormalized one with
 the cutoff for finite temperature, i.e., 
 $g_{\sigma+}(l=\ln \Lambda/T)$. 
Then the uniform spin susceptibility is given by
\begin{equation}
\chi_s(T) = \frac{\chi_0(T)}{ 1 - g_{\sigma+}(l) \chi_0(T) },
\label{eq:chi_s}
\end{equation}
   where $g_{\sigma+}(l)$ $[\equiv 2\pi v_F G_{\sigma+}(l)]$
   is obtained by solving the RG equation Eq.\ (\ref{eq:RG_gsigma+})
  or (\ref{eq:dG_sigma+/dl}).
This formula is valid at the temperature  above the energy scale of
  the gap $|m_t|$, since the coupling $g_{\sigma+}$ 
  which describe the fluctuation  of the Majorana triplet sector 
  [see Eq.\ (\ref{eq:Heff_spin})] is treated perturbatively.

\begin{figure}[b]
\includegraphics[width=6.cm]{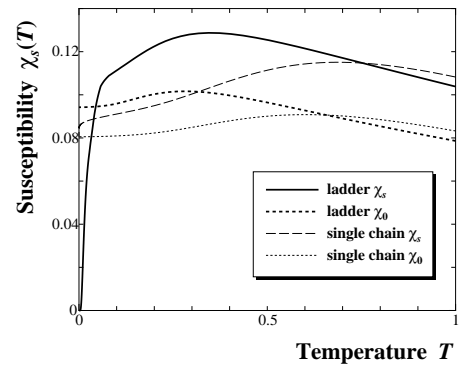}
\caption{
The temperature dependence of  several spin susceptibilities
  $\chi_s(T)$ and $\chi_0(T)$ for
$U/t=4$, $V_\parallel=V_\perp=0$, and $\delta=0.1$, with 
 $t_\parallel=t_\perp=1$, where
  $\chi_0$ denotes $\chi_s$ in the absence of interactions.
The Majorana triplet gap is $m_t\approx 0.03$.
For comparison, the corresponding susceptibilities of a single chain  
  are also shown.
}
\label{fig:chi-a}
\end{figure}

The overall temperature dependence of the uniform spin susceptibility 
  is shown in Fig.\ \ref{fig:chi-a}. 
At sufficiently high temperature, 
  the uniform susceptibility $\chi_s(T)$ exhibits
  behavior similar to  the one-dimensional
  susceptibility both in the presence and in the absence of interactions.
With decreasing temperature, $\chi_s(T)$ and $\chi_0(T)$ for the ladder 
  system become larger than those of a single chain due to the enhancement 
  of the density of states by interchain hopping.
However, $\chi_s(T)$ shows activation behavior 
below $|m_t|$.
In order to comprehend such low-temperature behavior, 
 we estimate the spin susceptibility based on the effective theory 
  for the spin mode, i.e., 
  the Majorana-fermion theory. 
Below the energy scale of the gap $|m_t|$, we could ignore 
  the fluctuation effects due to $g_{\sigma+}$ in Eq.\ (\ref{eq:Heff_spin}),
 which  would merely 
  yield the mass renormalization given in Eq.\ (\ref{eq:mt_tilde}).
Thus we can use the effective theory given by Eq.\
 (\ref{eq:Majorana_Hamiltonian}), although
there remain some discussions on $g_{\sigma+}$. \cite{Damle}
By using Eq.\ (\ref{eq:spin_current}) and
after
a straightforward calculation, we obtain the uniform susceptibility
as 
\begin{equation}
\chi_s(T)=\chi_{\mathrm{uni}}(q_\parallel,0,0)|_{q_\parallel\to 0}
= \frac{1}{8\pi T} \int
dk \,\, \mathrm{sech}^2 \frac{\varepsilon_k^t}{2T},
\label{eq:chis_Maj}
\end{equation}
  where $\varepsilon_k^t= (v_F^2 k^2 + m_t^2)^{1/2}$.
For low $T$ $(\ll |m_t|)$, Eq.\ (\ref{eq:chis_Maj}) is rewritten as
\begin{equation}
 \chi_s(T) \approx \frac{1}{v_F} \sqrt{\frac{|m_t|}{2\pi T}} e^{-|m_t|/T}.
\label{eq:chis_lowT}
\end{equation}
Thus we obtain the exponential decay of  the spin 
  susceptibility in the doped Hubbard ladder, which 
  is the same low-temperature asymptotics as in the undoped 
  Heisenberg ladder. \cite{Troyer,Kishine}

\begin{figure}[t]
\includegraphics[width=6.cm]{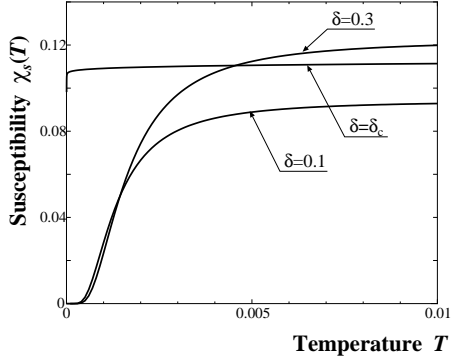}
\caption{
The temperature dependence of the spin susceptibility for
$U/t=3$, $V_\parallel=V_\perp=0.7$
with 
 $t_\parallel=t_\perp=1$.
The critical value of the doping is given by $\delta_c \approx 0.18$.
}
\label{fig:chi-b}
\end{figure}

The susceptibility at the QCP exhibits quite different behavior 
  from that of $V=0$.
Figure \ref{fig:chi-b} shows the ladder $\chi_s(T)$ at low temperature 
 for the doping near the QCP. 
Except for the QCP, the susceptibility exhibits 
  activation behavior at low temperature.
On the QCP, the  susceptibility shows paramagnetic temperature dependence
  for the whole temperature region due to the absence of the gap $m_t$.

\subsection{NMR relaxation rate}

In this section, we calculate the temperature dependence of the
NMR relaxation rate $T_1^{-1}$.
We use the general formula of the NMR relaxation rate
\cite{Moriya}
\begin{equation}
T_1^{-1}=
\frac{T}{2L}
\sum_{\bm{q}}  \frac{\chi''(\bm{q},\omega_0)}{\omega_0},
\label{T1general}
\end{equation}
  where $\chi''$ is the imaginary part of the magnetic susceptibility
  and $\omega_0$ is the nuclear resonance frequency having a small energy
  of the order of millikelvins.
Here we do not write the coefficient of the RHS of 
  Eq.\ (\ref{T1general}) by neglecting 
  the momentum dependence of the hyperfine coupling constant, 
since the hyperfine coupling 
  for the ladder-$^{63}$Cu site in Sr$_{14}$Cu$_{24}$O$_{41}$ 
  originates mainly in the on-site hyperfine interaction and thus 
  depends weakly on the momentum $\bm{q}$.\cite{Imai}
Since the interaction process between
  electrons  near Fermi points is considered to
  play an important role for weak coupling, 
   the integral over $q_\parallel$ in Eq.\ (\ref{T1general})
  can safely be split into two parts,
\begin{equation}
T_1^{-1}= (T_{1}^{-1})_{\mathrm{uni}} + (T_{1}^{-1})_{\mathrm{stag}},
\end{equation}
where $(T_{1}^{-1})_{\mathrm{uni}}$ and $(T_{1}^{-1})_{\mathrm{stag}}$
are the contributions from 
  small $q_\parallel$ and large $q_\parallel$, respectively.
Thus $(T_{1}^{-1})_{\mathrm{uni}}$ denotes
 the uniform contribution to  
  the NMR relaxation rate, while 
$(T_{1}^{-1})_{\mathrm{stag}}$ is
 the antiferromagnetic staggered contribution.
The explicit formulas are given by
\begin{subequations}
\begin{eqnarray}
&& (T_{1}^{-1})_{\mathrm{uni}} =
\frac{T}{4\pi} \sum_{q_\perp} \int_{q_\parallel\approx 0} dq_\parallel \, 
\frac{\chi''_{\mathrm{uni}}(\bm{q},\omega_0)}{\omega_0},
\label{eq:T1uni} \\
&& (T_{1}^{-1})_{\mathrm{stag}} =
\frac{T}{4\pi} 
\sum_{q_\perp}\int_{q_\parallel\approx \pm 2k_F} dq_\parallel \, 
\frac{\chi''(\bm{q},\omega_0)}{\omega_0}. \qquad
\label{eq:T1AF}%
\end{eqnarray}%
\end{subequations}
These two contributions $(T_{1}^{-1})_{\mathrm{uni}}$
 and $(T_{1}^{-1})_{\mathrm{stag}}$ show different temperature dependence.
It has been discussed which contribution becomes dominant 
 in Heisenberg ladder systems,
  \cite{Kishine,Ivanov1999,Sandvik1996,Naef} 
  but the problem is still controversial.
We treat these two contributions separately in the following.

\subsubsection{Uniform part: Contribution from small $q_\parallel$ }

Now we examine the uniform contribution $(T_{1}^{-1})_{\mathrm{uni}}$,
  by using the RPA calculation combined with the RG method at high
  temperature and by performing a direct calculation in terms of
  the low-energy effective theory at low temperature. 

First we focus on the high-temperature behavior of
$(T_1^{-1})_{\mathrm{uni}}$.
In order to obtain $(T_1^{-1})_{\mathrm{uni}}$ from Eq.\
(\ref{eq:T1uni}),  we  calculate the imaginary part of the
susceptibility $\chi''_{\mathrm{uni}}$.
In the noninteracting case, 
one easily finds from Eq.\ (\ref{eq:chi0}) that the imaginary part
of the noninteracting susceptibility becomes
\begin{subequations}
\begin{eqnarray}
\chi_0''(q_\parallel,0,\omega)  &=&
\frac{1}{4}
\sum_p pq_\parallel \, \delta(\omega-p v_F q_\parallel)
\label{eq:chi0imag}
\\
\chi_0''(q_\parallel,\pi,\omega)&=&
\frac{1}{8}\sum_{p,\zeta} (pq_\parallel + 2\zeta \lambda) \,
 \delta (\omega-pv_F q_\parallel - 2\zeta v_F \lambda),
\nonumber \\
\label{eq:chi0imag_pi}
\end{eqnarray}%
\end{subequations}
  where 
$\chi_0''(q_\parallel,q_\perp,\omega)
  \equiv \sum_{p=\pm} \mathrm{Im} \chi_0^p (q_\parallel,q_\perp,\omega)$ and
we have neglected the curvature of the dispersion and used the linearized
dispersion 
  $\varepsilon(k_\parallel,k_\perp) \to v_F (pk_\parallel-k_{F,k_\perp})$ 
 in Eq.\ (\ref{eq:chi0}).
Since $\chi_0''(\bm{q},\omega)$ is proportional to $\omega$ due to the
  $\delta$ function, the integral in  
  Eq.\ (\ref{eq:T1uni}) is determined by the contribution being linear
  in $\chi_0''$ in $\chi_{\mathrm{uni}}(q_\parallel,0,\omega)$    
  [Eq.\ (\ref{eq:chi_uni_0})].
Further, by neglecting the $q_\parallel$ and $\omega$ dependence 
  of $\mathrm{Re} \chi^p_0(q_\parallel,0,\omega)$ 
  (which is a nonsingular quantity) in Eq.\ (\ref{eq:chi_uni_0}),
the imaginary part of $\chi_{\mathrm{uni}}(q_\parallel,0,\omega)$ 
for small $q_\parallel$ and small $\omega$ is given by
\begin{eqnarray}
\chi''_{\mathrm{uni}}(q_\parallel,0,\omega) 
&\approx&
\frac{\chi''_0(q_\parallel,0,\omega)}
{ \left[   1-  g_{\sigma+} \chi'_0 (0,0,0) \right]^2 }
\nonumber \\
&\approx&
\frac{1}{4}
\frac{\chi_s^2(T)}{\chi_0^2(T)} \sum_p \, pq_\parallel \, 
\delta(\omega-pv_Fq_\parallel), \quad\quad
\label{eq:chiimag}
\end{eqnarray}
   where 
  $\chi'_0(q_\parallel,q_\perp,\omega)
   \equiv \sum_{p=\pm} \mathrm{Re}  \chi_0^p(q_\parallel,q_\perp,\omega)$
  and  $\chi_0(T)\equiv \chi'(0,0,0)$.
In the second equality of Eq.\ (\ref{eq:chiimag}), we have used 
   Eqs.\ (\ref{eq:chi_s}) and (\ref{eq:chi0imag}).
On the other hand, there is no correction to the $q_\perp=\pi$
  component,
  i.e., 
 $\chi''_{\mathrm{uni}}(q_\parallel,\pi,\omega)
  =\chi''_0(q_\parallel,\pi,\omega)$.
By inserting Eqs.\ (\ref{eq:chiimag}) and (\ref{eq:chi0imag_pi})
   into Eq.\ (\ref{eq:T1uni}), the uniform contribution
$(T_{1}^{-1})_{\mathrm{uni}}$ at $T\gg m_s,|m_t|$ is given 
by
\begin{equation}
(T_{1}^{-1})_{\mathrm{uni}}
=
\frac{T}{8\pi v_F^2} \,  \frac{\chi_s^2(T)}{\chi_0^2(T)}
+ \frac{T}{8\pi v_F^2}
.
\label{eq:T1a}
\end{equation}
The first (second) term in the RHS of Eq.\ (\ref{eq:T1a}) 
  comes from processes 
  with momentum transfer $q_\perp=0$ ($q_\perp=\pi$).
The contribution with $q_\perp=\pi$, which takes the same
  form as in the noninteracting case, 
  shows the Korringa law $T_1^{-1}\propto T$, while
the contribution with $q_\perp=0$ is enhanced by  the spin fluctuations
  yielding a  factor
  $\chi_s^2(T)/\chi_0^2 = 1/[1-g_{\sigma+}(T) \chi_0(T)]^2$.
The contribution of $q_\perp=0$ shows a relation 
  between the uniform part of the relaxation
  rate and the uniform spin susceptibility $\chi_s(T)$.

Next we calculate $(1/T_1)_{\mathrm{uni}}$ at low temperature 
  by using the low-energy effective spin Hamiltonian 
  (\ref{eq:Majorana_Hamiltonian}) which is valid at $T\ll m_{\rho-}$. 
By using Eqs.\ (\ref{eq:Majorana_Hamiltonian}) and (\ref{eq:spin_current}),
we obtain
\begin{widetext}
\begin{eqnarray}
(T_{1}^{-1})_{\mathrm{uni}} &=&
\frac{1}{16\pi v_F^2} \int_{|m_t|}^\infty d\varepsilon \,
  \frac{\varepsilon^2+m_t^2}
       {\sqrt{(\varepsilon^2 - m_t^2)[(\varepsilon + \omega_0)^2 -m_t^2]}} \,\,
\mathrm{sech}^2 \frac{\varepsilon}{2T}
\nonumber \\ && {}
+ \frac{1}{16\pi v_F^2} \int_{\max(|m_t|,m_s)}^\infty d\varepsilon \,
  \frac{\varepsilon^2+m_t m_s}
       {\sqrt{(\varepsilon^2 - m_t^2)(\varepsilon^2 - m_s^2)}} \,\,
\mathrm{sech}^2 \frac{\varepsilon}{2T},
\label{eq:T1_uni_maj_integral}
\end{eqnarray}
\end{widetext}
  where $\varepsilon_k^z= (v_F^2 k^2 + m_z^2)^{1/2}$ with $z=t,s$
  and we have set $\omega_0 \to 0$ in the second term of the RHS.
The first term in the RHS of Eq.\ (\ref{eq:T1_uni_maj_integral}) 
 is a contribution from the processes 
with momentum transfer $q_\perp=0$, which is given by the 
 Majorana triplet-triplet bubble in the diagram.\cite{Kishine}
On the other hand, the second term is a contribution from processes with
  $q_\perp=\pi$ and corresponds to the Majorana triplet-singlet bubble.
In the limit of high temperature,  
  Eq.\ (\ref{eq:T1_uni_maj_integral}) reduces to 
 $(T_1^{-1})_{\mathrm{uni}}=T/8\pi v_F^2+ T/8\pi v_F^2$ corresponding 
 to the Korringa law.
The enhancement factor in the first term of the RHS of Eq.\ (\ref{eq:T1a})
is not reproduced since the spin fluctuation has been neglected in 
Eq.\ (\ref{eq:T1_uni_maj_integral}).
At low temperature ($\omega_0\ll T \ll |m_t|$), the most dominant term reads
\begin{eqnarray}
(T_{1}^{-1})_{\mathrm{uni}}
\!\! &\approx& \!\! 
\frac{1}{4\pi v_F^2} 
\int_{|m_t|}^\infty d\varepsilon 
\frac{(\varepsilon^2+m_t^2) \,\, e^{-\varepsilon/T}}
      {\sqrt{(\varepsilon^2 - m_t^2)[(\varepsilon + \omega_0)^2 -m_t^2]}} 
\nonumber \\ 
\!\! &\approx& \!\! 
\frac{|m_t|}{4\pi v_F^2} e^{-|m_t|/T}
 K_0\left(\frac{\omega_0}{2T}\right)
\nonumber \\
\!\!&\approx& \!\! 
\frac{|m_t|}{4\pi v_F^2} e^{-|m_t|/T}
\left[
  \ln \left(\frac{4T}{\omega_0}\right) - \gamma \,
\right] ,
\label{eq:T1_uni_maj}
\end{eqnarray}
  where $K_0(z)$ is the modified Bessel function 
  and $\gamma$ is Euler's constant.
 Equation (\ref{eq:T1_uni_maj}) 
  is equal to the formula obtained in the undoped two-leg 
  Heisenberg ladder\cite{Troyer,Citro} and in 
  the spin-1 Haldane spin chain. \cite{Sagi}

\begin{figure}[t]
\includegraphics[width=6.cm]{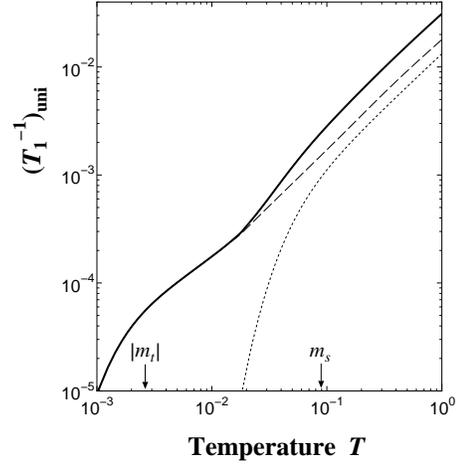}
\caption{
The temperature dependence of the uniform contribution of the 
 NMR relaxation rate,
with $U/t=3$, $V_\parallel/t=V_\perp/t=0.7$, $t\equiv 
 t_\parallel=t_\perp=1$,  and $\delta=0.1$. 
The dashed and dotted curves denote the $q_\perp=0$ and $\pi$ 
  contributions, respectively, and
  the solid curve denotes the total $(T_1^{-1})_{\mathrm{uni}}$.
}
\label{fig:t1}
\end{figure}

The overall temperature dependence of $(T_1^{-1})_{\mathrm{uni}}$ can be
  obtained by the interpolation between 
  Eqs.\ (\ref{eq:T1a}) and (\ref{eq:T1_uni_maj_integral}). 
We show the temperature dependence of 
$(T_1^{-1})_{\mathrm{uni}}$ in the CDW+PDW phase 
in Fig.\ \ref{fig:t1}.
In the SC$d$ phase,  $(T_1^{-1})_{\mathrm{uni}}$ shows behavior qualitatively 
  similar  to Fig.\ \ref{fig:t1}.
The dashed curve represents the $q_\perp=0$ contribution
while the dotted curve represents the $q_\perp=\pi$ contribution.
The $q_\perp=\pi$ contribution exhibits activation behavior
  at temperatures below $m_s$, while the $q_\perp=0$
  contribution shows $T$-linear dependence at temperatures above $|m_t|$.
For $T \ll |m_t|$, the NMR relaxation rate $(T_1^{-1})_{\mathrm{uni}}$
is governed by the first term in Eq.\ (\ref{eq:T1_uni_maj_integral}) 
 and exhibits an
activation behavior given by Eq.\ (\ref{eq:T1_uni_maj}).
The component with $q_\perp=0$ is always larger than that with $q_\perp=\pi$ 
 in the whole temperature region, in contrast to the results
  in Ref.\ \onlinecite{Kishine}.
It is found that the total $(T_1^{-1})_{\mathrm{uni}}$ clearly exhibits 
  behavior consisting of two components and thus the result 
  cannot be fitted by a single activation energy.

\subsubsection{Staggered part: Contribution from large $q_\parallel$}

We calculate the antiferromagnetic contribution
$(T_{1}^{-1})_{\mathrm{stag}}$ given by Eq.\ (\ref{eq:T1AF}).
As is known in the single-chain case,\cite{Bourbonnais}
  both the spin and charge degrees of freedom contribute to
  $(T_1^{-1})_{\mathrm{stag}}$. 
The critical charge mode in our theory is the total sector
$\phi_{\rho+}$, whose Hamiltonian is given by 
\begin{equation}
\mathcal{H}_{\rho+}
=\frac{v_{\rho+}}{2\pi}
\left[
 \frac{1}{K_{\rho+}} (\partial_x \phi_{\rho+})^2 +
 K_{\rho+} (\partial_x \theta_{\rho+})^2
\right],
\label{eq:charge_Gaussian}
\end{equation}
where $v_{\rho+}=v_F[1-(g_{\rho+}/2\pi v_F)^2]^{1/2}$ and
$K_{\rho+}=[(2\pi v_F -g_{\rho+})/(2\pi v_F +g_{\rho+})]^{1/2}
\approx 1- G_{\rho+}$.
For simplicity, we will neglect the normalization of the velocity and
  set $v_{\rho+}\to v_F$.

First we study $(T_{1}^{-1})_{\mathrm{stag}}$ at high temperature,
  i.e.,  $T\gg m_{\rho-}$.
For the noninteracting case,  the real part $\chi_0'$ 
   becomes logarithmic singular, i.e., 
   $\chi'_0(2k_F+q_\parallel,\pi,\omega,T)\approx (4\pi v_F)^{-1}
  \ln(\Lambda/x)$ where $x=\mathrm{max}(|v_F q_\parallel|,|\omega|,T)$,
  while  the imaginary part   $\chi_0''$ is nonsingular.
The imaginary part of $\chi_0$ for  $|\omega| \ll T$ is given by
\begin{equation}
\chi_0''(2k_F+q_\parallel,\pi,\omega,T) 
\approx \frac{\omega}{32 v_F T} \, 
\mathrm{sech}^{2} \left(\frac{v_Fq_\parallel}{4T}\right).
\end{equation}
By using the fact $|\chi_0'|  \gg |\chi_0''|$,
 the imaginary part of  $\chi$ is given by 
  \cite{Bourbonnais} 
\begin{eqnarray}
&&
\chi''(2k_F+q_\parallel ,\pi,\omega,T)
\nonumber \\ && {}
=  \overline{\chi}_{\mathrm{stag}}(q_\parallel,\omega,T)  \,\,
\chi_0''(2k_F + q_\parallel,\pi,\omega, T), \qquad
\label{eq:relation}
\end{eqnarray}
where $\overline{\chi}$ is an auxiliary function associated with
  the real part of $\chi$:
\begin{equation}
\overline{\chi}_{\mathrm{stag}}(q_\parallel,\omega,T) \equiv
4\pi v_F \,
\frac{\partial \, \mathrm{Re} \, \chi(2k_F+q_\parallel,\pi,\omega,T)}
  {\partial \ln (\Lambda/x)},
\label{eq:auxiliary}
\end{equation}
 with $x=\mathrm{max}(|\omega|,|v_F q_\parallel|,T)$.
The staggered part of the NMR relaxation rate can be obtained 
  by inserting Eq.\ (\ref{eq:relation}) into Eq.\ (\ref{eq:T1AF}).
Since $\chi''_0(2k_F + q_\parallel,\pi,\omega,T)$ 
  has a sharp peak around $q_\parallel=0$ with the exponential 
  decay for large $q_\parallel$, 
  we rewrite $\overline{\chi}_{\mathrm{stag}}(q_\parallel,\omega,T)$
  as  $\overline{\chi}_{\mathrm{stag}}(T)$ in Eq.\ (\ref{eq:relation})
  by   neglecting the $q_\parallel$ and $\omega$ dependence 
  in $\overline{\chi}_{\mathrm{stag}}(q_\parallel,\omega,T)$.
Then the staggered part
   $(T_{1}^{-1})_{\mathrm{stag}}$ is given by
\begin{equation}
(T_{1}^{-1})_{\mathrm{stag}}
=
\frac{T}{8\pi v_F^2}\, \overline{\chi}_{\mathrm{stag}} (T).
\label{eq:T1b}
\end{equation}
In order to examine $\overline{\chi}_{\mathrm{stag}}(T)$,
  we calculate the correlation function for $\bm{n}_-(x)$
   [Eq.\ (\ref{eq:spin_op})] which denotes the spin operator 
   with momentum transfer
  $q_\parallel \approx 2k_F$ and $q_\perp=\pi$. 
By applying the RG method to 
   the spin-spin correlation function 
 $R(x,\tau) =
 \langle T_\tau \, n^\alpha_-(x,\tau) \, n^\alpha_-(0,0) \rangle$,
   the RG equation for  $\overline{\chi}_{\mathrm{stag}}$ 
 is obtained as
\begin{eqnarray}
\frac{d}{dl} \ln \overline{\chi}_\mathrm{stag}(l)
&=&
 \frac{1}{2} G_{\rho+} - \frac{1}{2} G_{\rho-}
    + \frac{1}{2}  G_{\sigma+} -\frac{1}{2} G_{\sigma-}
\nonumber \\ && {} \quad
    + \frac{1}{2} G_{\overline{c-},st}
    + \frac{1}{2} G_{\overline{c-},ss}  
\label{eq:RG_chi_bar}
\end{eqnarray}
 where $l=\ln(\Lambda/T)$ and 
   $\overline{\chi}_\mathrm{stag}(0)=1$.
To obtain the temperature dependence of
$(T_1^{-1})_{\mathrm{stag}}$ from Eq.\ (\ref{eq:T1b}), we first solve 
 the RG equations  (\ref{eq:RG}) for the coupling 
 constants, and next substitute those into 
 Eq.\ (\ref{eq:RG_chi_bar}).
We note that,
for the noninteracting case, the auxiliary function is given by
  $\overline{\chi}_\mathrm{stag}(l)=1$
  and then  
  $(T_{1}^{-1})_{\mathrm{stag}}= T/8\pi v_F^2$, which has the same
  temperature dependence 
  as the uniform part [Eq.\ (\ref{eq:T1a})] in the noninteracting limit.

At temperature below the charge gap $m_{\rho-}$, i.e.,
for $l>l_{\rho-}$, the field $\theta_{\rho-}$ can be replaced by its average
value $\langle \theta_{\rho-} \rangle$. Without losing generality, 
 we can assume that $\theta_{\rho-}$ is locked at 
 $\langle\theta_{\rho-}\rangle=\pi I$ where $I$ is integer.
Then the spin operators are given by
\begin{subequations}
\begin{eqnarray}
n^x_-(x)
\!\! &\propto& \!\!
 \cos \tilde\phi_{\rho+} \sin \theta_{\sigma+} \cos \phi_{\sigma-}  ,
\\
n^y_-(x)
\!\! &\propto& \!\!
 \cos \tilde\phi_{\rho+} \cos \theta_{\sigma+} \cos \phi_{\sigma-} ,
\\
n^z_-(x)
\!\! &\propto& \!\!
 \cos \tilde\phi_{\rho+} \cos \phi_{\sigma+} \sin \theta_{\sigma-} ,
\end{eqnarray}%
\label{eq:spin_stagg_eff}%
\end{subequations}
where $\tilde{\phi}_{\rho+}$ is defined just after Eq.\ (\ref{eq:n-_boson}).
The RG equation for  $\overline{\chi}_{\mathrm{stag}}(T)$ is given by
\begin{equation}
\frac{d}{dl} \ln \overline{\chi}_\mathrm{stag}(l)
=
\frac{1}{2} +
 \frac{1}{2} G_{\rho+}
    + \frac{1}{2}  G_{\sigma+} -\frac{1}{2} G_{\sigma-}
    + M_{t}
    + M_{s}  ,
\label{eq:RG_chi_bar2}
\end{equation}
where the coupling constants are calculated
  from Eq.\ (\ref{eq:dG_spin}).
Thus the NMR relaxation rate $(T_1^{-1})_{\mathrm{stag}}$
 in the region of $\mathrm{max}(|m_t|,m_s) \ll T \ll m_{\rho-}$ is 
 given by Eqs.\ (\ref{eq:T1b}) and (\ref{eq:RG_chi_bar2}).
However, in the present system, we find $m_s \approx m_{\rho-}$,
  and then this asymptotic behavior would not be realized.

Next we calculate the antiferromagnetic contribution to the relaxation
  rate at low temperature
  by using the effective spin Hamiltonian (\ref{eq:Majorana_Hamiltonian}).
In this case, the staggered components 
  of the spin operators (\ref{eq:spin_stagg_eff})
  are rewritten as
\begin{subequations}
\begin{eqnarray}
n^x_-(x)
\!\! &\approx& \!\!
 \cos \tilde\phi_{\rho+} \left(\sigma_1 \mu_2 \mu_3 \right) \mu_4,
\\
n^y_-(x)
\!\! &\approx& \!\!
 \cos \tilde\phi_{\rho+} \left(\mu_1 \sigma_2 \mu_3 \right) \mu_4,
\\
n^z_-(x)
\!\! &\approx& \!\!
 \cos \tilde\phi_{\rho+} \left(\mu_1 \mu_2 \sigma_3 \right) \mu_4,
\end{eqnarray}%
\label{eq:n_Ising}%
\end{subequations}
where the Klein factors have been omitted.
We follow the calculation performed in Refs.\
  \onlinecite{Ivanov1999,Schulz1986}, and \onlinecite{Sachdev1994}.
If the spin-spin correlation function at zero temperature exhibits 
  power-law behavior with exponent $\eta$, i.e.,
  $\langle n_-(x) n_-(0) \rangle \propto x^{-\eta}$, 
  it can be shown  that 
  the temperature dependence of the NMR relaxation rate is 
  given by $(T_1^{-1})_{\mathrm{stag}} \propto T^{\eta -1}$, 
by using the conformal mapping technique.  \cite{Sachdev1994}
If  $|m_t| , |m_s| \ll T \ll m_{\rho-}$ is realized, 
  the exponent is $\eta=\frac{1}{2}K_{\rho+}+1$ and
 then we obtain
\begin{eqnarray}
(T_{1}^{-1})_{\mathrm{stag}} \propto T^{K_{\rho+}/2}.
\label{eq:t1sta_a}
\end{eqnarray}
This behavior, however, would not be observed since $m_s\approx m_{\rho-}$.
For $|m_t| \ll T \ll |m_s|$ in the
SC$d$ or CDW+PDW phase,
 we can replace $\mu_4$ by its average value 
  $\langle \mu_4 \rangle \neq 0$ in Eq.\ (\ref{eq:n_Ising}) and then
  the exponent of the spin-spin correlation function becomes
  $\eta=\frac{1}{2}K_{\rho+} + \frac{3}{4}$.
Thus we have
\begin{eqnarray}
(T_{1}^{-1})_{\mathrm{stag}} \propto T^{-1/4+K_{\rho+}/2}.
\label{eq:t1sta_b}
\end{eqnarray}
In the limit of low temperature ($T\ll |m_t|, |m_s|, m_{\rho-}$), 
the relaxation rate exhibits thermally activated behavior. 
In the SC$d$ phase ($m_t>0$), we obtain (see the Appendix)
\begin{widetext}
\begin{eqnarray}
(T_{1}^{-1})_{\mathrm{stag}}
&\propto&
\sum_{\epsilon=\pm}
\int_{-\infty}^{\infty} \frac{d\omega}{2\pi}
\cosh \left(i\epsilon \frac{\pi}{2}\eta + \frac{\omega}{2T} \right)
\left(\frac{2\pi T}{v_F}\right)^{\eta-1}
B\left(\frac{\eta}{2}-i\epsilon \frac{\omega}{2\pi T}, 1-\eta\right)
\nonumber \\ && {} \times
\int_{-\infty}^{\infty} 
\frac{d\theta_1}{2\pi} \frac{d\theta_2}{2\pi} \frac{d\theta_3}{2\pi} \,
\frac{2\pi \delta 
  \biglb( E(\theta_1) -E(\theta_2)+E(\theta_3)+\omega\bigrb)}
 { 8 \cosh [E(\theta_1)/2T] \, \cosh [E(\theta_2)/2T] \,
     \cosh [E(\theta_3)/2T] } 
  \, \coth^2 \left(\frac{\theta_1-\theta_2}{2}\right) ,
\label{eq:t1sta_c0}
\end{eqnarray}
\end{widetext}
where $E(\theta)=|m_t| \cosh \theta$ is the rapidity representation of 
  dispersion, 
  $B(x,y)$ is the beta function, and  $\eta =K_{\rho+}/2$.
In the low temperature limit $T\ll m_t$, the staggered part of the 
  NMR relaxation rate in the SC$d$ state is given by (see the Appendix)
\begin{equation}
(T_{1}^{-1})_{\mathrm{stag}} \propto
T^{1+K_{\rho+}/2} \, \exp(-2 m_t /T),
\label{eq:t1sta_c}
\end{equation}
showing the activation behavior with a gap $2m_t$. 
On the other hand, in the CDW+PDW state ($m_t<0$),
  by considering two magnon process, 
  the NMR relaxation rate would be given by
  (see the Appendix)
\begin{equation}
(T_{1}^{-1})_{\mathrm{stag}} \propto
  T^{K_{\rho+}/2} \, \exp(-|m_t| /T) \, \ln T .
\label{eq:t1sta_c2}
\end{equation}

\begin{figure}[t]
\includegraphics[width=6.cm]{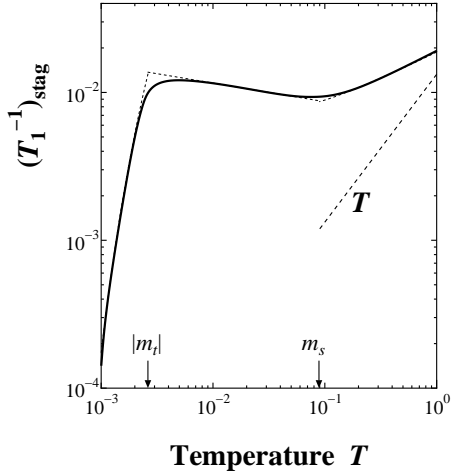}
\caption{
The temperature dependence of the NMR relaxation rate
$(T_1^{-1})_{\mathrm{stag}}$
for $U/t=3$, $V_\parallel/t=V_\perp/t=0.7$, $t\equiv 
 t_\parallel=t_\perp=1$,  and $\delta=0.1$. 
}
\label{fig:t1stag}
\end{figure}

In Fig.\ \ref{fig:t1stag}, we show the staggered part of the NMR  relaxation
rate, $(T_1^{-1})_{\mathrm{stag}}$ at low temperature,
  which is calculated
  based on Eqs.\ (\ref{eq:t1sta_a}), (\ref{eq:t1sta_b}), and
  (\ref{eq:t1sta_c}).
The enhancement of $(T_1^{-1})_{\mathrm{stag}}$ 
  in the interval region
  of $|m_t|<T<m_s$, which is similar to that of single chain,
  originates in the antiferromagnetic fluctuations.
The rapid decrease of  $(T_1^{-1})_{\mathrm{stag}}$ at temperature of 
$T<|m_t|$  shares the common feature with $(T_1^{-1})_{\mathrm{uni}}$ 
  and $\chi_s$.
We note that the relative magnitude of $(T_1^{-1})_{\mathrm{uni}}$ 
  and 
  $(T_1^{-1})_{\mathrm{stag}}$ depends on those of their hyperfine
  couplings.

\subsection{Quantum critical behavior}

Finally we focus on the temperature dependence of $\chi_s$, 
   $(T_1^{-1})_{\mathrm{uni}}$, and  $(T_1^{-1})_{\mathrm{stag}}$,
  just above the QCP, i.e., $m_t=0$.
Since the $g_{\sigma+}$ term becomes  marginally irrelevant 
  in the effective theory (\ref{eq:Heff_spin}),
 we first examine the scaling of $g_{\sigma+}$ which 
  gives rise to logarithmic corrections for the physical quantities.

In the effective spin Hamiltonian (\ref{eq:Heff_spin}),
  we consider the temperature region lower than the gap in the Majorana
  singlet sector.
In this case, we can integrate out the $\xi^4$ degrees of freedom
  and then we can rewrite 
  $m_t^0 -i g_{\sigma-}\langle \xi_+^4 \xi_-^4 \rangle \to m_t^0$.
Further we can set $m_t^0=0$ on the critical point.
Then only the $g_{\sigma+}$ term remains where 
  the RG equation  is given by
\begin{equation}
\frac{d}{dl} G_{\sigma+}(l) =
-G_{\sigma+}^2(l),
\label{eq:g_sigma+_QCP}
\end{equation}
  and $g_{\sigma +}(l)=(2\pi v_F) \, G_{\sigma+}(l) $.
By solving Eq.\ (\ref{eq:g_sigma+_QCP}), we have
\begin{equation}
g_{\sigma+}(l)= \frac{g_{\sigma+}}{1+(g_{\sigma+}/2\pi v_F)(l-l_s)},
\label{eq:g_rg}
\end{equation}
where the initial value is given by $g_{\sigma+}(l_s)=g_{\sigma+}
  \equiv (2\pi v_F) \,G_{\sigma+}(l_s)$.
The quantity $l_s$ corresponds to the 
  scale of the gap in the Majorana singlet excitation,  
 $m_s \approx \Lambda e^{-l_s}$.

From Eq.\ (\ref{eq:g_sigma+_QCP}), 
  the character of the phase transition is determined by the 
  sign of the initial value of $G_{\sigma +}$. 
\cite{Tsuchiizu2002b,Fradkin2002,Starykh}
When the initial value is 
 given by $G_{\sigma +}(l_s)>0$,
the coupling constant $G_{\sigma+}(l)$ decreases 
to zero under renormalization, and then
    becomes marginally irrelevant.
In this case, the effective theory in the low-temperature limit 
 leads to the \textit{noninteracting} massless Majorana fermion, 
 and thus the system exhibits a quantum critical behavior.
On the other hand, for the initial value being negative, i.e., 
  $G_{\sigma +}(l_s)<0$,
  the coupling constant $G_{\sigma +}(l)$ becomes marginally relevant 
  due to its divergence at $l_t=2\pi v_F/|g_{\sigma+}|$.
In this case, the effective theory does not 
  give a quantum critical behavior 
  due to  a mass gap $m_t \approx \Lambda e^{-2\pi v_F/|g_{\sigma +}|}$,
  even if the bare mass $m_t^0$ reduces to zero.
From the numerical calculation of Eqs.\ (\ref{eq:RG}) and (\ref{eq:dG_spin}), 
  we have confirmed that the coupling constant $G_{\sigma+}$ at $l=l_s$
  is positive within our choice of repulsive interactions,   
 and that the present ladder system corresponds to the former case, i.e., 
  the system exhibits the quantum critical behavior.
If the sign of $G_{\sigma +}(l_t)$ could be changed by 
   another type of interaction in the 
  microscopic Hamiltonian,\cite{Tsuchiizu2002b,Fradkin2002}
    the first-order transition would be obtained 
   instead of the QCP within the present framework.

In the following we examine  the temperature dependence of  $\chi_s(T)$,
   $(T_1^{-1})_{\mathrm{uni}}$, and  $(T_1^{-1})_{\mathrm{stag}}$,
  separately by using Eq.\ (\ref{eq:g_rg}) 
  with $l=\ln(\Lambda/T)$ where  $\Lambda$ is a
  high-energy cutoff of the order of the bandwidth.
We note that at low temperature, i.e., for large $l$, 
  the renormalized coupling
  constant shows  $g_{\sigma+}(l) \approx 2\pi v_F/ \ln(\Lambda/T)$.

\subsubsection{The uniform spin susceptibility $\chi_s(T)$}

We can use the formula (\ref{eq:chi_s})
  for calculating  the uniform spin susceptibility
since $m_t=0$ at the QCP.
By inserting Eq.\ (\ref{eq:g_rg}) into Eq.\ (\ref{eq:chi_s}),
the temperature dependence of $\chi_s(T)$  is obtained with
 the low-temperature asymptotics,
\begin{equation}
\chi_s(T) \approx\frac{1}{2\pi v_F}
 \left[1 + \frac{1}{\ln(\Lambda/T)}\right].
\label{eq:chis_QCP}
\end{equation}
Figure \ref{fig:chi-b} shows the temperature dependence of $\chi_s(T)$
  where $\delta=\delta_c$ corresponds to the QCP.
Equation (\ref{eq:chis_QCP}) is compared with 
  the spin susceptibility for the
  $S=\frac{1}{2}$ Heisenberg single chain given by
  \cite{Eggert}
\begin{equation}
\chi_s^{\mathrm{1D}}(T) \approx \frac{1}{2\pi v}
 \left[1 + \frac{1}{2} \, \frac{1}{\ln(\Lambda/T)}\right],
\label{eq:chis_QCP1D}
\end{equation}
  where $v=\pi J/2$ and $J$ is the exchange interaction.
The results (\ref{eq:chis_QCP}) and (\ref{eq:chis_QCP1D}) are
  consistent with the susceptibility in
  the  SU(2)$_k$ WZNW critical theory with marginally irrelevant operators,
  which gives the logarithmic correction  $\{1+ k/[2 \ln(\Lambda/T)]\}$ 
  where $k$ is the level of SU(2) algebra. \cite{Babujian}
Note that $k=2$ for a two-leg ladder,
  while $k=1$ for the $S=\frac{1}{2}$ single chain.

\begin{figure}[t]
\includegraphics[width=6.cm]{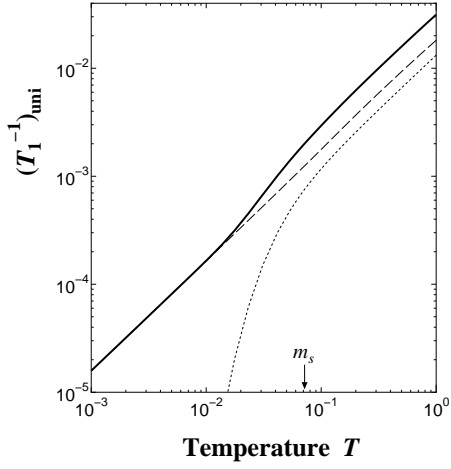}
\caption{
The temperature dependence of the NMR relaxation rate
with $U/t=3$, $V_\parallel/t=V_\perp/t=0.7$, $t\equiv 
 t_\parallel=t_\perp=1$,  and $\delta=\delta_c \approx 0.18$.
The power-law behavior is retained even in the limit of low temperature
  due to the vanishing of 
  the Majorana triplet mass, i.e., $m_t=0$. 
The dashed and dotted curves correspond to the
  contributions with  $q_\perp=0$ and $\pi$, respectively, while
the solid curve denotes the total $(T_1^{-1})_{\mathrm{uni}}$.
}
\label{fig:t1QCP}
\end{figure}

\subsubsection{The NMR relaxation rate: 
    $(T_1^{-1})_{\mathit{uni}}$ and  $(T_1^{-1})_{\mathit{stag}}$}

The temperature dependence of $(T_1^{-1})_{\mathrm{uni}}$  can be obtained
  by inserting Eq.\ (\ref{eq:chis_QCP}) into Eq.\ (\ref{eq:T1a}).
The second term of the RHS of Eq.\ (\ref{eq:T1a}) 
  can be discarded since this would show the exponential decay 
  at temperature below $m_s$ as seen from the dotted curve of
  Fig.\ \ref{fig:t1QCP}.
For the low-temperature limit, we have
\begin{equation}
(T_{1}^{-1})_{\mathrm{uni}} \approx
\frac{T}{8\pi v_F^2} \left[1 + \frac{2}{\ln(\Lambda/T)}\right].
\label{eq:T1u_QCP}
\end{equation}
The overall temperature dependence of $(T_{1}^{-1})_{\mathrm{uni}}$
  at the QCP is shown in Fig.\ \ref{fig:t1QCP}.
For the single chain with $S=\frac{1}{2}$,  we have 
\begin{eqnarray}
(T_{1}^{-1})_{\mathrm{uni}}^{\mathrm{1D}}
 &\approx& \frac{T}{4\pi v_F^2} \,
   \frac{[\chi_s^{\mathrm{1D}}(T)]^2}{\chi_0^2(T)}
\nonumber \\
&\approx&
\frac{T}{4\pi v_F^2} \left[1 + \frac{1}{\ln(\Lambda/T)}\right],
\label{eq:T1u_QCP1D}
\end{eqnarray}
where the first equality is obtained in Ref.\  \onlinecite{Bourbonnais}.

The logarithmic correction to
the staggered part $(T_1^{-1})_{\mathrm{stag}}$ can be obtained as follows.
If one neglects the marginally irrelevant $g_{\sigma+}$,
  the staggered part  $(T_1^{-1})_{\mathrm{stag}}$
  is given by Eq.\ (\ref{eq:t1sta_b}).
In order to retain the renormalization effect of $g_{\sigma+}$,
  we use Eq.\ (\ref{eq:T1b}).
In a way similar to the derivation of Eq.\ (\ref{eq:g_sigma+_QCP}), 
   the scaling equation of the auxiliary field
   $\overline{\chi}_{\mathrm{stag}}(T)$
  on the QCP at temperature below $m_s$ is given by
\begin{equation}
\frac{d}{dl} \ln \overline{\chi}_\mathrm{stag}(l)
= \frac{3}{4} + \frac{1}{2} G_{\rho+} +\frac{1}{2} G_{\sigma+}(l),
\label{eq:RG_chi_bar3}
\end{equation}
  where $G_{\rho+}$ is independent of $l$ and the
  factor $3/4$ in the RHS is determined from the 
  scaling dimension of the spin operator Eq.\ (\ref{eq:n_Ising}),
  i.e., $(2-2\mathrm{dim} [\cos \tilde{\phi}_{\rho+}] 
        - 2 \mathrm{dim} [\sigma_1 \mu_2 \mu_3] ) =3/4$.
By solving Eq.\ (\ref{eq:RG_chi_bar3}) with $K_\rho=1-G_{\rho+}$, we have 
\begin{equation}
\overline{\chi}_\mathrm{stag}(T) =
\left( \frac{T}{\Lambda}\right)^{-5/4+K_{\rho+}/2}
\sqrt{1 + \frac{g_{\sigma+}}{2\pi v_F} \ln \left(\frac{\Lambda}{T}\right)}.
\label{eq:RG_chi_bar3_sol}
\end{equation}
From Eqs.\ (\ref{eq:T1b}) and (\ref{eq:RG_chi_bar3_sol}),
  we obtain the low-temperature asymptotics of
  $(T_1^{-1})_{\mathrm{stag}}$ is obtained as
\begin{equation}
 (T_1^{-1})_{\mathrm{stag}} \propto  
T^{-1/4+K_{\rho+}/2} \sqrt{\ln \left(\frac{\Lambda}{T}\right)}.
\label{eq:T1stag_QCP}
\end{equation}
This result is compared with the staggered component of $T_1^{-1}$
  for the single chain, which is given by
\begin{equation}
 (T_1^{-1})_{\mathrm{stag}}^{\mathrm{1D}}
\propto 
T^{K_{\rho}}
\sqrt{\ln \left(\frac{\Lambda}{T}\right)}.
\label{eq:T1stag_QCP1D}
\end{equation}
We note that, in the insulating state
 $(K_\rho \to 0)$, 
  Eq.\ (\ref{eq:T1stag_QCP1D})  is reduced to
 $ (T_1^{-1})_{\mathrm{stag}}^{\mathrm{1D}} \to 
  \sqrt{\ln (\Lambda/T)}$ 
   reproducing the result of the $S=\frac{1}{2}$ Heisenberg spin
  chain, \cite{Barzykin}
  while Eq.\ (\ref{eq:T1stag_QCP})
  leads to 
 $ (T_1^{-1})_{\mathrm{stag}} \to T^{-1/4} \sqrt{\ln (\Lambda/T)}$ which 
  is consistent with the result obtained in Ref.\ \onlinecite{Ivanov1999} 
  except for the logarithmic correction.

\section{Conclusions and Discussion}\label{sec:conclusions}

In the present paper, we have examined the ground-state phase diagram 
 and the temperature dependence of the susceptibility and 
  the NMR relaxation rate
  for the extended two-leg Hubbard model away from half filling,
  by using the weak-coupling bosonization method.
In the ground state, we have clarified the competition between
   the SC$d$ state and the CDW+PDW state
  and have shown the quantum critical behavior close to
 the transition point where the SC$d$ state changes into
  the CDW+PDW state with increasing the nearest-neighbor repulsion 
   and/or decreasing doping rate.
At finite temperature, 
  the magnetic response exhibits characteristic property coming from
 two modes of spin excitations.
Especially on the quantum critical point, we found that
the spin susceptibility shows paramagnetic temperature dependence
  with logarithmic corrections
and the NMR relaxation rate exhibits anomalous power-law behavior.

Here we discuss the commensurability effect due to the umklapp
  scattering which would play an important role close to half filling. 
At half filling, the umklapp scattering is given by
 \cite{Tsuchiizu2002b}
\begin{eqnarray}
\mathcal{H}_{\mathrm{umklapp}}
\!\!&=&\!\!
   \frac{1}{4}
   \sum_{p,\sigma}{\sum_{\zeta_i=\pm}}'
[  g_{3\parallel}^{\epsilon\bar\epsilon} \,
      \psi_{p,\sigma,\zeta_1}^\dagger \,
      \psi_{p,\sigma,\zeta_2}^\dagger \,
      \psi_{-p,\sigma,\zeta_4}^{} \,
      \psi_{-p,\sigma,\zeta_3}^{}
\nonumber \\&& {} 
  + g_{3\perp}^{\epsilon\bar\epsilon} \,
      \psi_{p,\sigma,\zeta_1}^\dagger \,
      \psi_{p,-\sigma,\zeta_2}^\dagger \,
      \psi_{-p,-\sigma,\zeta_4}^{} \,
      \psi_{-p,\sigma,\zeta_3}^{} ], 
\nonumber \\
\label{eq:Hint_umklapp}
\end{eqnarray} 
where
$g_{3\parallel}^{\epsilon\bar\epsilon}
=  l_\epsilon V_\perp + m_{3,\epsilon} V_\parallel$
 and
$g_{3\perp}^{\epsilon\bar\epsilon}
= (U + l_\epsilon  V_\perp + m_{3,\epsilon} V_\parallel)$
   with the numerical factors
   $l_\pm = \pm 1$, 
   $m_{3,+}=-1$, and $m_{3,-}=-2$.
In terms of bosonic fields,
  Eq.\ (\ref{eq:Hint_umklapp}) is rewritten as
\begin{eqnarray}
\mathcal{H}_{\mathrm{umklapp}}
&\!\!=\!\!& 
  \frac{1}{2\pi^2 a^2} [ 
    g_{c+, \overline{c-}} \,
    \cos 2 \phi_{\rho+} \,
    \cos 2 \theta_{\rho-} 
\nonumber \\ && {}
+   g_{c+, s+} \,
    \cos 2 \phi_{\rho+} \,
    \cos 2 \phi_{\sigma+}   
  \nonumber \\
&& {}
+   g_{c+, s-} \, 
    \cos 2 \phi_{\rho+} \,
    \cos 2 \phi_{\sigma-}   
  \nonumber \\
&& {}
+   g_{c+, \overline{s-}} \,
    \cos 2 \phi_{\rho+} \,\,
    \cos 2 \theta_{\sigma-}   ],
\label{eq:Hintum}
\end{eqnarray}
where the coupling constant are
$g_{c+,\overline{c-}} =  - g_{3\perp}^{-+}$,
$g_{c+,s+} = - g_{3\parallel}^{+-} + g_{3\parallel}^{--}$,
$g_{c+,s-} = - g_{3\perp}^{+-}$, and
$g_{c+, \overline{s-}} = + g_{3\perp}^{--}$.
For the rung-singlet state at half filling, 
  the renormalized coupling constants in 
Eq.\ (\ref{eq:Hintum}) are given by
$ g_{c+, \overline{c-}}^*<0$,
  $ g_{c+, s+}^*<0$,
  $ g_{c+, s-}^* <0$, and
  $ g_{c+, \overline{s-}}^*=0$.
Even in the presence of a few holes, the
  renormalized umklapp scattering would remain  finite 
  unless at extremely low energy scale.
Then here we fix the amplitude of the umklapp scattering 
  and discuss the effect of the finite doping.
In terms of phase variable, the particle number operator is given by
\begin{eqnarray}
N&=&
\sum_{j,l,\sigma}
\left(
 c_{j,l,\sigma}^\dagger \, c_{j,l,\sigma}^{}
 - \frac{1}{2}
\right)
\nonumber\\
&=&
\frac{2}{\pi} [ \phi_{\rho+}(\infty) - \phi_{\rho+}(-\infty) ].
\label{eq:N}
\end{eqnarray}
From Eq.\ (\ref{eq:N}), the injection of a single electron or hole
 corresponds to the formation of the $\pi/2$ soliton or 
  antisoliton
  in  the $\phi_{\rho+}$ mode.
In order to avoid the increase of energy,  
 the $\pi/2$ soliton in the $\phi_{\rho+}$ mode is always accompanied by the 
  $\pi/2$ solitons in the 
  $\phi_{\sigma+}$, $\theta_{\rho-}$, and $\phi_{\sigma-}$ modes.
This fact implies that 
   the $\pi/2$ soliton in the $\phi_{\rho+}$ mode 
   involves the appearance of local spin at the same rung.
On the other hand, the $\pi$ soliton in the $\phi_{\rho+}$ mode,
  which corresponds to $N=2$,
 is not accompanied by solitons in the spin and other modes.
Thus, if a $\pi/2$ soliton in the $\phi_{\rho+}$ mode is 
  created in the system, 
  the free spin appears at the same rung in the bulk rung-singlet state
  and the spin-charge separation does not take place.
This picture would connect with the strong-coupling one 
  in the sense that holes can destroy  spin singlets 
  in the two-leg ladder systems. \cite{Dagotto}

Finally we compare the present results with the experimental ones on
  the two-leg ladder
  compounds Sr$_{14-x}$Ca$_x$Cu$_{24}$O$_{41}$, which have
  the characteristic features of the spin-gapped normal state and 
  the superconducting state.
For $x=12$ and under a pressure of 3.5 GPa, the NMR measurements
  show two excitation modes above the SC state, where 
  $T_1^{-1}$ decreases rapidly at higher temperature and
  $T$-linear dependence is found at lower temperature.
  \cite{Fujiwara}
This result resembles the present result of Figs.\ \ref{fig:t1} 
  and \ref{fig:t1QCP}.
The decrease for $T>m_s$ comes from the  formation of the spin
  gap in the spin-singlet excitations while the linear dependence
  for $|m_t|<T<m_s$ appears due to the gapless mode with the freedom of 
  spin-triplet excitations.
Such an interval region is enlarged close to the QCP since $m_t\to 0$
  at the QCP, as seen from Fig.\ \ref{gap}.
From the NMR shifts, it is shown that the uniform magnetic
susceptibility decreases slowly for $30 < T < 200~\mathrm{K}$
 and stays constant for $T<30~\mathrm{K}$.
The former resembles Fig.\ \ref{fig:chi-a}
  where the slow decrease below $T/t<0.3$ is due to the band effect.
The latter would correspond to Fig.\ \ref{fig:chi-b} with $\delta=0.1$,
  where $\chi_s$ is almost independent of temperature for 
  $0.07<T/t<0.01$.
Thus the present scenario of two spin excitations could be relevant to
  experiments although the present approach is based
  on weak-coupling theory.

\acknowledgments

M.T.\ thanks  A.\ Furusaki, E.\ Orignac, and O.A.\ Starykh 
   for valuable discussions.
This work was supported by a Grant-in-Aid for
   Scientific Research on Priority Areas of Molecular Conductors
   (Grant No.\ 15073213) 
   from the Ministry of Education, Culture, Sports, Science and Technology,
   Japan.

\appendix

\section{Staggered part of the NMR relaxation rate 
  in the low-temperature limit}

In this appendix, we derive the staggered part of the NMR relaxation rate
  $(T_{1}^{-1})_{\mathrm{stag}}$  
  in the low-temperature limit 
  [Eqs.\ (\ref{eq:t1sta_c0})$-$(\ref{eq:t1sta_c2})],
  based on the Majorana-fermion description of the effective theory.

In the temperature region $T\ll (|m_s|, m_{\rho-})$, 
 the staggered component of the
  spin operator [Eq.\ (\ref{eq:n_Ising})] is
  rewritten as
\begin{subequations}
\begin{eqnarray}
n^x_- &\propto& \cos \tilde\phi_{\rho+} \, (\sigma_1 \, \mu_2 \, \mu_3),
\\
n^y_- &\propto& \cos \tilde\phi_{\rho+} \, (\mu_1 \, \sigma_2 \, \mu_3), 
\\
n^z_- &\propto& \cos \tilde\phi_{\rho+} \, (\mu_1 \, \mu_2 \, \sigma_3),
\end{eqnarray}
\end{subequations}
where $\tilde\phi_{\rho+}=\phi_{\rho+}-\pi\delta x$.
In terms of these operators, the NMR relaxation rate is given by
\begin{eqnarray}
(T_{1}^{-1})_{\mathrm{stag}}
&\propto& \int_{-\infty}^{\infty} dt \,\, S(t) ,
\label{eq:ap_T1}
\end{eqnarray}
where $S(t)=\langle n_-^\alpha (\mbox{$x=0$},t) \,\, n_-^\alpha(0,0)\rangle$
  is the local correlation function at finite temperature.
We will estimate this correlation function by using the 
  effective Hamiltonian 
  (\ref{eq:QI}) and (\ref{eq:charge_Gaussian}).
Since the charge and spin degrees of freedom are decoupled,
  the correlation function can be rewritten as
\begin{equation}
S(t) = S_{\rho+}(t) \, S_{\mathrm{Ising}}(t) ,
\label{eq:ap_S}
\end{equation}
  where 
\begin{subequations}
\begin{eqnarray}
S_{\rho+}(t) &\equiv&
    \langle \cos \phi_{\rho+}(0,t) \, \cos \phi_{\rho+}(0,0) \rangle ,
\\
S_{\mathrm{Ising}}(t) &\equiv& 
   \langle \mu(0,t) \, \mu(0,0) \rangle^2 \,
     \langle \sigma(0,t) \, \sigma(0,0) \rangle.
\end{eqnarray}
\end{subequations}
One can easily find that
   all the correlation functions
  $\langle n_-^\alpha(0,t) n_-^\alpha(0,0) \rangle$ for $\alpha=x,y,z$ become
  identical, since the system has spin-rotational symmetry.

The  local correlation function for the charge fields is given by
  \cite{Schulz1986}
\begin{eqnarray}
S_{\rho+}(t) &=&
  \frac{1}{2} \, e^{-i(\pi/2)\eta \, \mathrm{sgn \,} (t)}
  \left[ \frac{\pi T a/v }{\sinh (\pi T |t|)} \right]^\eta
\nonumber \\
&=&
\int_{-\infty}^{\infty} \frac{d\omega}{2\pi} e^{-i\omega t}
  \sum_{\epsilon=\pm} \frac{a}{2v} 
  e^{-i(\pi/2)\eta}\left( \frac{2\pi Ta}{v} \right)^{\eta-1}
\nonumber \\ && {} \quad \times
  B\left(\frac{\eta}{2}-i\epsilon \frac{\omega}{2\pi T}, 1-\eta\right) ,
\label{eq:ap_rho}
\end{eqnarray}
where $\eta=K_{\rho+}/2$ and $B(x,y)$ is the beta function
  $B(x,y)=\Gamma(x) \, \Gamma(y)/\Gamma(x+y)$.
In the second equality, we have performed the Fourier transformation.
  \cite{Schulz1986}

The correlation function for the Ising fields at finite temperature 
  can be calculated following Ref.\ \onlinecite{LeClair}.
The asymptotic behavior of the Ising correlation function depends on 
  whether the system is in the ordered phase or in the disordered phase,
  i.e., depends on the sign of the mass $m_t$.

In the SC$d$ phase ($m_t>0$), the Ising systems $\sigma_i$ $(i=1,2,3)$
  are in the disordered phase; thus $\mu_i$ has a nonzero expectation value.
In this case, the dominant contribution in the low-temperature limit is 
  \cite{Ivanov1999}
\begin{eqnarray}
&&\left. 
S_{\mathrm{Ising}}(t)
\right|_{m_t>0}
\propto
\sum_{\epsilon,\epsilon_3=\pm}
\int_{-\infty}^{\infty} 
\frac{d\theta_1}{2\pi}\frac{d\theta_2}{2\pi}\frac{d\theta_3}{2\pi}
\nonumber \\ && {} \qquad \times
f_\epsilon(\theta_1) \, f_{-\epsilon}(\theta_2) \, f_{\epsilon_3}(\theta_3) \, 
\coth^2\left(\frac{\theta_1-\theta_2}{2}\right)
\nonumber \\ && {} \qquad \times
e^{-i[\epsilon E(\theta_1)
      -\epsilon E(\theta_2)+\epsilon_3 E(\theta_3)]t},
\label{eq:ap_Ising}
\end{eqnarray}
where  $E(\theta)=|m_t| \cosh \theta$ and
   $f_\epsilon(\theta)=[1+e^{-\epsilon E(\theta)/T}]^{-1}$.
By inserting Eqs.\ (\ref{eq:ap_rho}) and (\ref{eq:ap_Ising})
into Eq. (\ref{eq:ap_S}), and inserting it into 
  Eq.\ (\ref{eq:ap_T1}), 
   we obtain Eq.\ (\ref{eq:t1sta_c0}).

In the CDW+PDW phase ($m_t<0$), the Ising systems are now 
  in the ordered phase; thus $\sigma_i$ has a nonzero expectation value. 
In this case the dominant contribution at low temperature reads 
\begin{eqnarray}
&&
\left. 
S_{\mathrm{Ising}}(t) 
\right|_{m_t<0}
\propto
\sum_{\epsilon=\pm}
\int_{-\infty}^{\infty} 
\frac{d\theta_1}{2\pi}\frac{d\theta_2}{2\pi}
\nonumber \\ && {} \qquad \times
f_\epsilon(\theta_1) \, f_{-\epsilon}(\theta_2) \,
e^{-i\epsilon [E(\theta_1) - E(\theta_2)]t}.
\end{eqnarray}
Then the NMR relaxation rate for $m_t<0$ is given by
\begin{eqnarray}
&& 
(T_{1}^{-1})_{\mathrm{stag}}  \propto
\sum_{\epsilon=\pm}
\int_{-\infty}^{\infty} \frac{d\omega}{2\pi}
\cosh \left(i\epsilon \frac{\pi}{2}\eta + \frac{\omega}{2T} \right)
\nonumber \\ && {} \quad \times
\left(\frac{2\pi T}{v_F}\right)^{\eta-1}
B\left(\frac{\eta}{2}-i\epsilon \frac{\omega}{2\pi T}, 1-\eta\right)
\nonumber \\ && {} \quad \times
\int_{-\infty}^{\infty} 
\frac{d\theta_1}{2\pi} \frac{d\theta_2}{2\pi} \,
\frac{2\pi \delta 
  \biglb( E(\theta_1) -E(\theta_2)+\omega\bigrb)}
 { 4 \cosh [E(\theta_1)/2T] \, \cosh [E(\theta_2)/2T] },
\nonumber \\
\label{eq:t1sta_c0cdw}
\end{eqnarray}

The low-temperature asymptotic forms of 
 $(T_{1}^{-1})_{\mathrm{stag}}$ [Eqs.\ (\ref{eq:t1sta_c}) and
  (\ref{eq:t1sta_c2})] are obtained as follows.
The $\omega$ integral in Eq.\ (\ref{eq:t1sta_c0})
  or Eq.\ (\ref{eq:t1sta_c0cdw}) is cut by the temperature $\pm T$,
  since by performing the summation with respect to 
  $\epsilon$ $(=\pm)$ in Eq.\ (\ref{eq:t1sta_c0})
  or Eq.\ (\ref{eq:t1sta_c0cdw}) one obtains  (for $|\omega|\gg T$)
\begin{eqnarray}
&&\sum_{\epsilon=\pm}
\cosh \left(i\epsilon \frac{\pi}{2}\eta + \frac{\omega}{2T} \right)
B\left(\frac{\eta}{2}-i\epsilon \frac{\omega}{2\pi T}, 1-\eta\right)
\nonumber \\ && \qquad \approx
\sin (\pi \eta) \, \Gamma(1-\eta) \,
\left( \frac{|\omega|}{2\pi T}\right)^{\eta-1} \, 
\exp\left( - \frac{|\omega|}{2T}\right).
\nonumber \\
\end{eqnarray}
This can easily be verified by using 
  the asymptotic form of the beta function:
\begin{equation}
B\left(\frac{\eta}{2}-i S, 1-\eta\right)
\approx \Gamma(1-\eta) \, (-iS)^{\eta-1} \quad \mbox{(for $S \to \pm \infty$)}
\end{equation}
Thus we have for $m_t>0$
\begin{eqnarray}
(T_{1}^{-1})_{\mathrm{stag}} 
&\propto&
T^\eta
\int_{-\infty}^{\infty} d\theta_1 d\theta_2 d\theta_3 \, 
 \, \coth^2 \left(\frac{\theta_1-\theta_2}{2}\right)
\nonumber \\ && {}  \times
 \delta \biglb( E(\theta_1) -E(\theta_2)+E(\theta_3)\bigrb)
\nonumber \\ && {} \times
e^{-[E(\theta_1) +E(\theta_2)+E(\theta_3)]/2T}
\nonumber \\ 
&\propto& T^{1+\eta} \exp(-2m_t/T) ,
\end{eqnarray}
which reproduces Eq.\ (\ref{eq:t1sta_c}).
For $m_t<0$ we have 
\begin{eqnarray}
(T_{1}^{-1})_{\mathrm{stag}} 
&\propto&
T^\eta
\int_{-\infty}^{\infty} d\theta_1 d\theta_2  \, 
 \delta \biglb( E(\theta_1) -E(\theta_2)\bigrb)
\nonumber \\ && {} \times
e^{-[E(\theta_1) +E(\theta_2)]/2T}
\nonumber \\ 
&\propto& T^{\eta} \exp(-|m_t|/T) \, \ln T, 
\end{eqnarray}
which reproduces Eq.\ (\ref{eq:t1sta_c2}).

\end{document}